\documentclass[usenatbib,usegraphicx]{mn2e}
\usepackage{lscape}

\title[Optical properties  of the NGC~5328 group of galaxies]
{Optical properties of the NGC~5328 group of galaxies{\thanks{Based on 
observations collected at European Southern Observatory, La Silla, Chile  
(Programme Nr. 65.P--252)}}}

\author[R.~Gr\"utzbauch  et al.] 
{R.~Gr\"utzbauch$^{1}$, F.~Annibali$^{2}$, A.~Bressan$^{2,3}$, P.~Focardi$^{4}$, B.~Kelm$^{4}$,
\newauthor R.~Rampazzo$^{3}$, W.W.~Zeilinger$^{1}$  \\
$^{1}$ Institut f\"ur Astronomie, Universit\"at Wien, T\"urkenschanzstra{\ss}e 17, A-1180 Wien, Austria\\
$^{2}$ SISSA, via Beirut 4, I-34014 Trieste, Italy\\
$^{3}$ INAF - Osservatorio Astronomico di Padova, Vicolo dell'Osservatorio 5, I-35122 Padova, Italy \\
$^{4}$ Dipartimento di Astronomia,Universit\`a di Bologna, Via Berti Pichat 6, I-40127 Bologna, Italy}
\date{Accepted. Received; in original form }

\pagerange{\pageref{firstpage}--\pageref{lastpage}}
\pubyear{2005}

\begin{document}

\maketitle

\label{firstpage}

\begin{abstract}
We present the results of a photometric and spectroscopic study of seven members
of the NGC~5328 group of galaxies, a chain of galaxies spanning over 200~kpc
(H$_{0}$=70 km~s$^{-1}$Mpc$^{-1}$). We analyze the galaxy structure and study
the emission line properties of the group members looking for signatures
of star formation and AGN activity. We finally attempt to infer, from
the modeling of line-strength indices, the stellar population ages of the
early-type members. We investigate also the presence of a dwarf galaxy
population associated with the bright members.
  
The group is composed of a large fraction of early-type galaxies
including NGC~5328 and NGC 5330 two ``bona fide'' ellipticals at the
center of the group. In both galaxies no recent star formation
episodes are detected by the H$\beta$ vs. MgFe line-strength indices
of these galaxies. 2MASX J13524838-2829584 has extremely boxy
isophotes which are believed to be connected to a merging event: line
strength indices suggest that this object probably had a recent star
formation episode. A warped disc component emerges from the model
subtracted image of 2MASX J13530016-2827061 which is interpreted as a
signature of an ongoing interaction with the rest of the group.

Ongoing star formation and nuclear activity is present in the
projected outskirts of the group. The two early-type galaxies 2MASX
J13523852-2830444 and 2MASX J13525393-2831421 show spectral signatures
of star formation while a Seyfert 2 type nuclear activity is detected
in MCG~--5--33--29.
\end{abstract}


\section{Introduction}

Recent investigations tend to map galaxy properties in the cluster
outskirts and beyond, at local densities typical of galaxy groups and
where a pre-processing of the subsequent cluster galaxies could take
place \citep{Lewis2002,BB2004,mih04}. The dynamically relaxed centers
of clusters are surrounded by infall regions in which galaxies are
bound to the cluster but not in an equilibrium configuration. The mass
of such an infall region could be of the order 20 - 120\% of the
virial mass, showing that clusters are still forming \citep{rin03}.
The galaxy population in that region is not pristine, since
significant preprocessing in group-like environments are expected
\citep{gil05}.  Groups of galaxies, of different richness and density,
are also typical galaxy aggregates of low density environments. Since
they contain a substantial fraction of the mass of the Universe, it is
of crucial importance to shed light about the galaxy evolution in such
an environments.

The low velocity dispersion in groups leads to slow encounters between
member galaxies and is predicted to trigger star formation episodes
\citep[see e.g.][]{ken96} and AGN activity. Nearby groups do not show,
however, firm evidence for galaxy activity enhancement
\citep{ver98,coz00,mai03,kel04,tan03}. The interaction processes are
not only altering the overall characteristics of the involved
galaxies, they also could affect the group galaxy population (e.g via
galaxy-galaxy merging and/or formation of tidal dwarf galaxies
\citep[see e.g.][and references therein]{Duc04} and of the IGM.

Recent X-ray studies have shown that a significant amount of a diffuse
emission could be present in galaxy groups \citep[see
e.g.]{M93,M96,pon96,mul00, mul03,gom03} and even in physical pairs
composed of an elliptical and a spiral member
\citep{ram98,HC99,TR01}. However it is still unclear whether massive
groups, whose dominant galaxy is a spiral, do exist
\citep{mul00,kelfoc04}.

Evolutionary phenomena in poor structures are furthermore of interest
in the context of a evolving galaxy groups since candidate ``fossil
groups'', i.e. the debris of a pristine group, are found at low and
intermediate redshifts \citep{mul99,jon03}.  The evolutionary scenario
of isolated galaxy groups raises then fundamental questions: what is
the final stage of small galaxy structures in the field? Are they
stable and long-lived structures or does their evolution lead to the
formation of giant elliptical galaxies in the field traced by the
detected X--ray diffuse emission?

Several arguments indicate that photometric parameters of massive
ellipticals are not compatible with their formation being the result
of several major merging events diluted along the Hubble time
\citep{mez03}. Moreover, massive early-type galaxies in low density
environments appear on the average only 2 Gyr younger than their
counterparts in high density environments \citep{thom05}. Because a
considerable fraction of massive galaxy formation in low density
environments occurs at redshifts $1 < z < 2$, the end product of
groups observed out to $z \approx 1$ could not easily be a new massive
early-type galaxy.

In this paper we present the optical, photometric and spectroscopic
properties of the poor group dominated by the bright elliptical
galaxy NGC~5328. The group has a compact and elongated shape: galaxies
are basically aligned in a long chain spanning from Northeast to
Southwest. NGC~5328 is well centered within this galaxy chain.  This
group is located in the vicinity of the Abell~3574 cluster of
galaxies. Our photometric and spectroscopic investigation covers the
central part of the group including seven spectroscopically confirmed,
bright member galaxies distributed over a radius of about 100 kpc
(H$_{0}$=70 km~s$^{-1}$~Mpc$^{-1}$).  We also provide positions and
the basic photometric properties of dwarf galaxies, lying within our
frames, candidate members of the group.  

The paper is organized as follows. In Section~2 we summarize the
literature data about NGC 5328 group. Section~3 presents the
observations and the data reduction.  The detailed surface photometry,
the analysis of galaxy structures and a preliminary description of the
spectroscopic properties of each galaxy are given in
Section~4. Medium-resolution spectroscopy is used to establish group
memberships and to analyze the spectral energy distributions in order
to check possible features like induced activity and enhanced star
formation.  Line-strength indices which, transformed into the Lick-IDS
system, were modeled to infer the star formation history of the
bright galaxies in the group, are presented in Section~5. The group
properties are finally discussed in the context of the literature and
in particular of our previous works on this subject
\citep[][paper~I]{tan03} and \citet[][paper~II]{gru05}.

\section{The NGC 5328 group in the literature}

The NGC~5328 group of galaxies was first mentioned by \citet{kle69} as
the object nr.~28 in his catalogue of ``Groups and Clusters of
Southern Galaxies''.  The author described the structure as a ``group
of round or somewhat elongated galaxies with one spiral to NE''. In
the original description, Klemola~28 consists of 7 galaxies within an
area of 15\arcmin$\times$5\arcmin, although the member galaxies are
not named. Given the coordinates of Klemola 28 and the description of
the group structure, we presume that the members of Klemola~28
coincide with the bright galaxies of the present study.

\citet{gar93} associates NGC~5328 with the well studied Seyfert galaxy
IC~4329 and 21 other galaxies to the group LGG~357.  A quite similar
set of galaxies is given by \citet{giu00} in his NOG~725. None of the
galaxies considered in our study is part of the LGG or the NOG
sample. IC~4329 is located at a projected distance of about 2.5 Mpc
from NGC~5328 and belongs to the Abell~3574 cluster of galaxies, a
poor cluster consisting of 55 spectroscopically confirmed member
galaxies moving with a recession velocity of $cz = 4797$ km s$^{-1}$
and having a velocity dispersion of $\sigma$ = 793 km
s$^{-1}$ \citep{str99}.  In the red-shift space our group and the
cluster form a coherent structure with a virtually null difference in
their systemic velocity (v$_{cluster}$ -- v$_{NGC~5328}$ = 39 km
s$^{-1}$).  The projected separation between the NGC~5328
group and the cluster outskirts amounts to $\approx$
1$^{\circ}$20$^{\prime}$ (corresponding to 1.7~Mpc). Thus, 
although the group cannot be characterized
as a substructure of the cluster, the group and the cluster could be
potentially connected. \citet{ric84} presents a list of prominent
galaxies on the plates 444 and 445 of the ESO-SRC Survey, giving the
morphological classifications for all galaxies in our sample and
including redshifts for NGC~5328 and NGC~5330.

Concerning the properties of individual galaxies of our sample,
MCG~--5--33--29 is quoted in the ``Catalog of Southern Ringed
Galaxies'' \citep{but95} but only NGC~5328 has been studied in detail
so far. \citet{dec88} presented a detailed surface photometry,
indicating that NGC~5328 is a ``bona fide'' elliptical. Several values
for the central velocity dispersion are available, with an average of
303 km ~s$^{-1}$ ({\tt http://www-obs.univ-lyon1.fr/hypercat}), the
latest measurement by \citet{smi00} amounts to $\sigma$ = 314 $\pm$ 7
km ~s$^{-1}$ suggesting that it is a massive galaxy. NGC~5328 is a
relatively strong X-ray emitter.  \citet{beu99} report a value of
$\log L_X = 42.03 \pm 0.072 ~erg ~s^{-1}$ within a radius of
16.25\arcmin.  The morphology and the characteristics of the X-ray
emission are not discussed. \citet{mah01} presented a unified relation
between L$_X$ and $\sigma$ for 280 clusters and 57 galaxies of the
form L$_X \propto \sigma^m$ with a slope steepening from $m$=4.4 for
clusters to $m$=10.2 for individual galaxies. The relation for groups
is not that well defined and shows a larger scatter suggesting that
most of them have either not yet reached dynamical equilibrium or the
emission is produced by unresolved sources in the IGM. NGC~5328 is
included in the sample of individual galaxies and lies exactly on the
L$_X \propto \sigma^{10.2}$ relation. The discussion of the X-ray
properties of NGC~5328 as a group, and its evolutionary phase, then
awaits more detailed X-ray observations.

\begin{table}
\caption{Observing log of imaging}
\label{tab1}
\begin{tabular}{lcclccc} \hline
{(1)} & {(2)} & {(3)} & {(4)} & {(5)} & {(6)} & {(7)}\\ 
\hline
1 & 13$^h$52$^m$54$^s$ & --28$^{\circ}$29$'$05$''$ & 4$\times$60  & 1.10 & 
2$\times$300 & 1.05 \\
2 & 13$^h$52$^m$47$^s$ & --28$^{\circ}$31$'$36$''$ & 3$\times$30  & 1.00 & 
3$\times$300 & 0.96 \\
3 & 13$^h$53$^m$09$^s$ & --28$^{\circ}$27$'$18$''$ & 3$\times$60  & 1.13 & 
3$\times$300 & 1.08 \\
\hline
\end{tabular}

\medskip
{$^1$}{~field nr.};
{$^2$}{~$\alpha$ (J2000.0)};
{$^3$}{~$\delta$ (J2000.0)};
{$^4$}{~R-band exposure time [s]};
{$^5$}{~R-band seeing FWHM [arcsec]};
{$^6$}{~B-band exposure time [s]};
{$^7$}{~B-band seeing FWHM [arcsec]}.
\end{table}

\section{Observations and data reduction}

The observations were performed at the ESO 3.6m telescope in La Silla, Chile 
during the nights of 5$^{th}$ and 6$^{th}$ May 2000.
Details of the observations and the data reduction procedures are therefore
described in paper~II.

In order to cover the central part of the NGC~5328 group a mosaic of 3
images has been assembled. The observing log of imaging containing the
field coordinates, exposure time and seeing conditions is given in
Table~\ref{tab1}.  Long--slit medium--resolution spectra have been
obtained for 8 galaxies to obtain the spectral information of the
candidate member galaxies.  Table~\ref{tab2} shows the instrumental
set--up of the spectroscopic observations. The velocity dispersion
values given in Table \ref{tab3} were determined by performing
Gaussian fits to single spectral lines. The FWHM of the fitted
profiles were corrected for the instrumental dispersion, but the large
instrumental dispersion of $\sigma_{inst}$= 227 km s$^{-1}$ biases the
measurements toward higher values. The values of $\sigma$ in Table
\ref{tab3} can therefore only give a rough estimate of the galaxies'
true velocity dispersion.

\begin{table*}
\caption{Spectroscopic observations}
\label{tab2}
\begin{tabular}{ccclll} \hline
{field}   & {exp.~t.} & {sl.-wid.} & {P.A.}       & {object}  & {wav.~range} 
\\ 
  {nr.}    &      {[s]}    & {[\arcsec]} & {[$^\circ$]} &      {}       &      
{[{\AA}]} \\
\hline
 1 & 2$\times$1200 & 1.0$''$ & 132$^\circ$ & NGC~5328                      & 4070 -- 7470 \\
         &               &         &       & NGC~5330                      & 4070 -- 7470 \\
 2 & 2$\times$1200 & 1.0$''$ & 111$^\circ$ & 2MASX J13524838-2829584       & 4070 -- 7470 \\
         &               &         &       & 2MASX J13523852-2830444       & 4070 -- 7470 \\
 2 & 2$\times$1200 & 1.0$''$ & 10$^\circ$  & 2MASX J13525552-2834001 (B1)  & 4070 -- 7470 \\
         &               &         &       & 2MASX J13525393-2831421       & 4070 -- 7470 \\
 3 & 2$\times$1200 & 1.0$''$ & 115$^\circ$ & MCG~--5--33--29               & 4070 -- 7470 \\
         &               &         &       & 2MASX J13530016-2827061       & 4070 -- 7470 \\
\hline
\end{tabular}
\end{table*}

\begin{figure*}
\resizebox{16.0cm}{!}{
\includegraphics[]{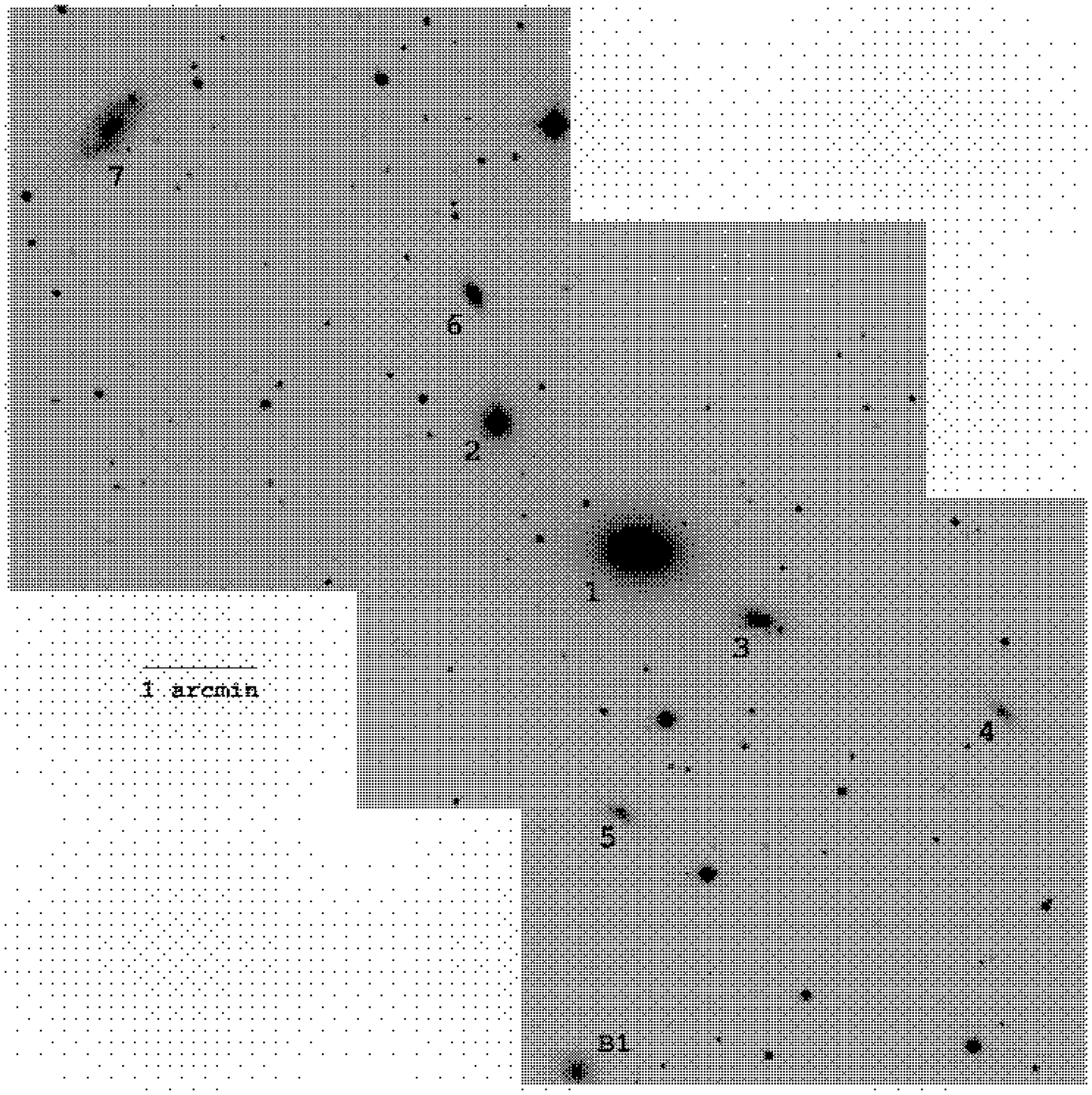}}
\caption{B-band mosaic image of the NGC~5328 group of galaxies. The group
members are numbered accordant to Table \ref{tab3}. The redshift of the 
galaxy labelled B1 has been measured and is not concordant with that of the group. 
\label{fig1}}
\end{figure*}

The mosaic image of the group shown in Figure~\ref{fig1} covers
$\approx$ 60 arcmin$^2$. The seven spectroscopically confirmed member
galaxies listed in Table~\ref{tab3} are indicated in the figure.  The
group structure has an elongated shape, six of seven member galaxies
are aligned in a chain spanning from the southwest to the
northeast. The bright elliptical galaxy NGC~5328 lies roughly in the
center of this galaxy chain. The mean projected separation of the
seven member galaxies amounts to $\approx$200 kpc.

\begin{table*}
\caption{Spectroscopic Results}
\label{tab3}
\begin{tabular}{lccllccc} \hline
{object}   & {$\alpha$} & {$\delta$} & {type}  & {V$_{hel}$}        & {D$^1$}  &   $\sigma_{0}$&   {id. Fig.1} \\ 
                &(J2000.0)  & (J2000.0)       &            & [km~s$^{-1}$]     &  [kpc]      &    [km~s$^{-1}$]              &      \\
\hline
NGC~5328        & 13$^h$52$^m$53.6$^s$ & --28$^{\circ}$29$'$16$''$ & E3                             &  4785 $\pm$ 131 & --& 303 $^{2}$   & 1 \\
NGC~5330        & 13$^h$52$^m$59.2$^s$ & --28$^{\circ}$28$'$15$''$ & E1                             & 4737 $\pm$ 60  & 35 & 292 $\pm$  60  &2 \\
2MASX J13524838-2829584       & 13$^h$52$^m$48.4$^s$ & --28$^{\circ}$29$'$58$''$ & S0     & 4566 $\pm$ 76  & 26 &  246 $\pm$ 118 &3 \\
2MASX J13523852-2830444       & 13$^h$52$^m$38.5$^s$ & --28$^{\circ}$30$'$45$''$ & S0     & 5452 $\pm$ 43  & 63 &                     & 4 \\
2MASX J13525393-2831421       & 13$^h$52$^m$53.9$^s$ & --28$^{\circ}$31$'$42$''$ & SB0   & 5153 $\pm$ 51  & 47 &                 &5 \\
2MASX J13530016-2827061       & 13$^h$53$^m$00.2$^s$ & --28$^{\circ}$27$'$06$''$ & S0     & 5243 $\pm$ 90  & 57 &  298 $\pm$ 127   &6 \\
MCG~--5--33--29 & 13$^h$53$^m$15.6$^s$ & --28$^{\circ}$25$'$33$''$ & SB(r)a                     & 4625 $\pm$ 164& 124&                 &7 \\
\hline
\end{tabular}

\medskip 
{$^1$}{~Projected distance from NGC~5328 (H$_0$ = 70 $km s^{-1}Mpc^{-1}$)}; 
{$^2$}{~The value of the velocity dispersion quoted  for NGC~5328
is taken from \citet{ramp05} since it has been used for the calibration of
line-strength indices to the Lick-IDS systems.}

\end{table*}

In order to extend our understanding of the group properties we
attempted the detection of candidate dwarf galaxies members of the
group in our B and R band exposures. The {\tt SourceExtractor}
\citep{ber96} was used to detect extended sources, i.e. having a {\it
stellaricity} parameter $\leq$ 0.5. The candidate sample was limited
in magnitude ($m{_R} \leq 21^{mag}$) and size ($a \geq 2$ arcsec),
furthermore, a colour restriction was applied in the attempt to
exclude background objects \citep[see e.g.][]{kho04}.  The trend with
fainter magnitudes toward bluer colours, reflecting the expected
lower metallicities of less massive galaxies, is present in our group
sample. For our faint galaxy sample, this trend is well defined in the
blue colour regime, while on the red side the contamination with
background objects becomes visible. A colour restriction of 0.7 $\leq$
(B--R) $\leq$ 1.9, representing generous limits to the early-type
sequence, was found to be likely to remove the vast majority of
background objects. The resulting sample of likely dwarf galaxy group
members consists of 68 objects with a mean colour of (B--R) = 1.32
$\pm$ 0.28 mag. The magnitudes and the (B--R) colour of this 68
extended sources are given in the Appendix, Tables~A1 and A2.  We did
not perform surface photometry on the dwarf galaxy candidates because
our observations are not deep enough to provide an accurate surface
photometry.

\section{Results}

The relevant spectroscopic and photometric results about the seven
member galaxies of the NGC~5328 group are summarized in
Table~\ref{tab3} and Table~\ref{tab4} respectively. We collect in
Table \ref{tab5} the results of a research (done using the {\tt NED})
about galaxies with known redshift within 1 Mpc from NGC~5328
representing possible members/neighbours of the group. These latter
are not uniformly distributed around the group, but are located within
two narrow cones perpendicular to the orientation of the chain.

In the sub-sections below we detail the photometric and spectroscopic results.

\subsection{Photometric results and individual notes}

Surface photometry of the bright group member galaxies has been
carried out with the ellipse fitting routine provided by the {\tt
STSDAS} package within IRAF\footnote{{\tt IRAF} is distributed by the
National Optical Astronomy Observatories, which are operated by the
Association of Universities for Research in Astronomy Inc., under
cooperative agreement with the National Science Foundation.} and with
the {\tt GALFIT} package \citep{pen02}. While the {\tt ELLIPSE} task
computes a Fourier expansion for each successive isophote
\citep{jed87}, {\tt GALFIT} was used to perform a bulge-disc
decomposition in the case of early-type galaxies and to determine the
parameters of a Sersic model fit to the galaxies bulge component. The
Sersic profile is the generalization of the de Vaucouleur's law with
$\mu(r) \sim r^{1/n}$, where the Sersic parameter $n$ is a free
parameter.  This profile is thus sensitive to structural differences
between different kinds of early type galaxies and providing a better
fit to real galaxy profiles. Only the surface brightness profile of
NGC~5328 was fitted by $n \sim 4$, whereas the other objects are
featuring a variety of Sersic parameters in the range $1 \leq n \leq
4$.

\begin{figure*}
\resizebox{17.0cm}{!}{
\includegraphics[]{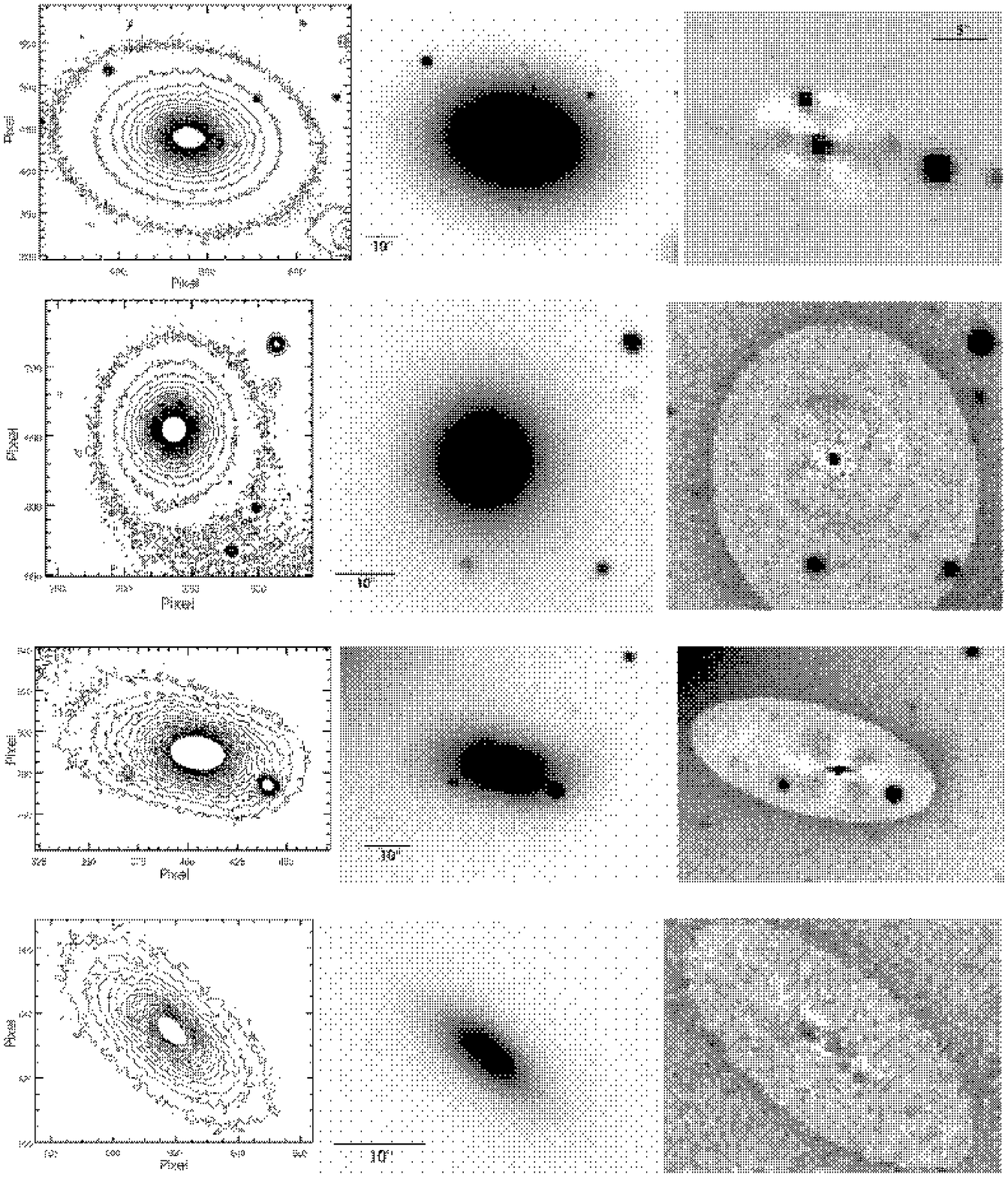}}
\caption{Analysis of fine structure in galaxies: isophote structure
(left panels), original image (mid panels) and model subtracted images
(right panels). From top to bottom: NGC~5328, NGC~5330, 2MASX
J13524838-2829584 and 2MASX J13523852-2830444.
\label{fig2}}
\end{figure*}

\begin{figure*}
\resizebox{17.0cm}{!}{
\includegraphics[]{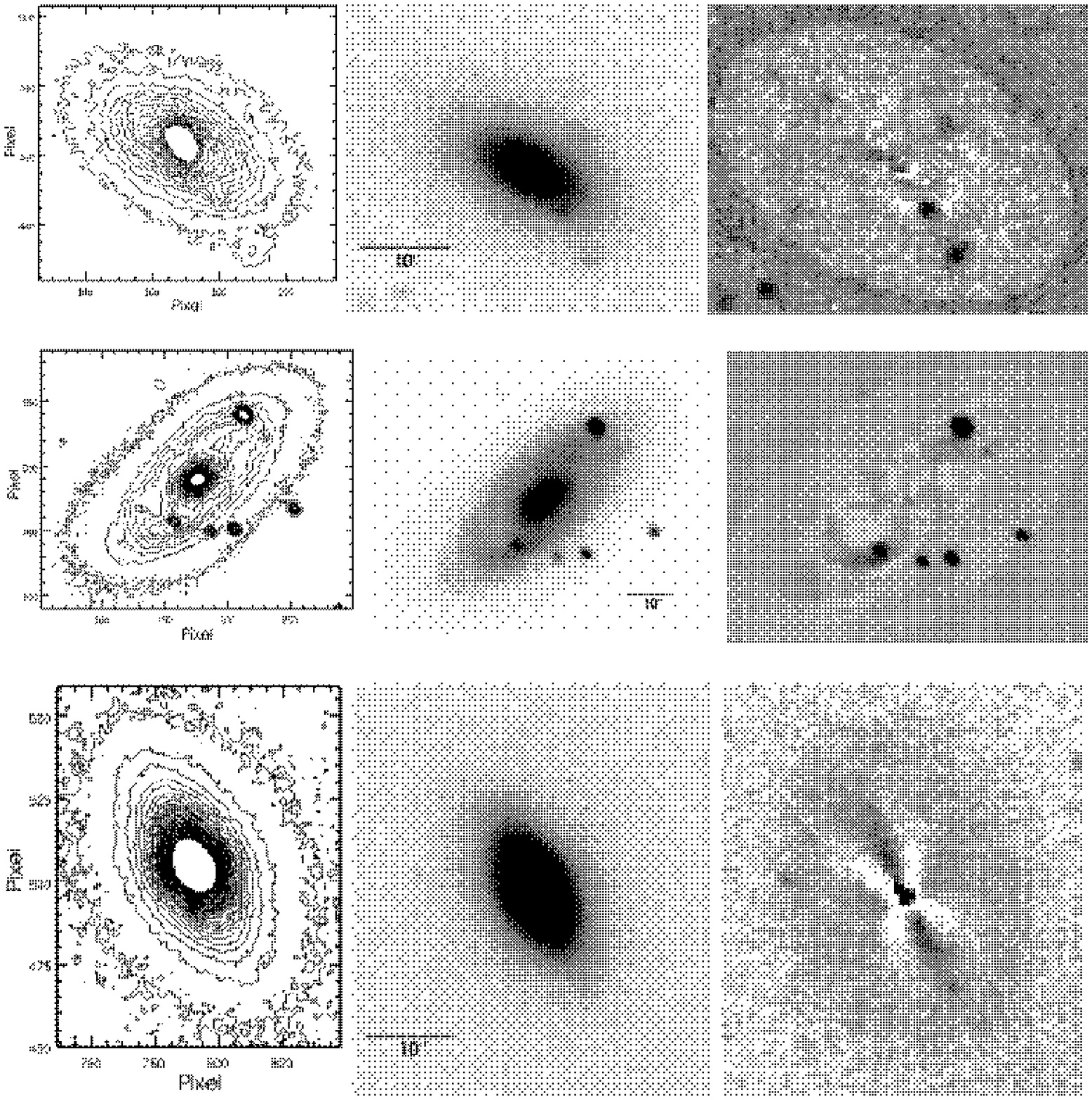}}
\caption{Analysis of fine structure in galaxies: isophote structure
(left panels), original image (mid panels) and model subtracted images
(right panels). From top to bottom: 2MASX J13525393-2831421,
MCG~--5--33--29 and 2MASX J13530016-2827061.
\label{fig3}}
\end{figure*}

\begin{figure*}
\resizebox{16cm}{!}{
\includegraphics[]{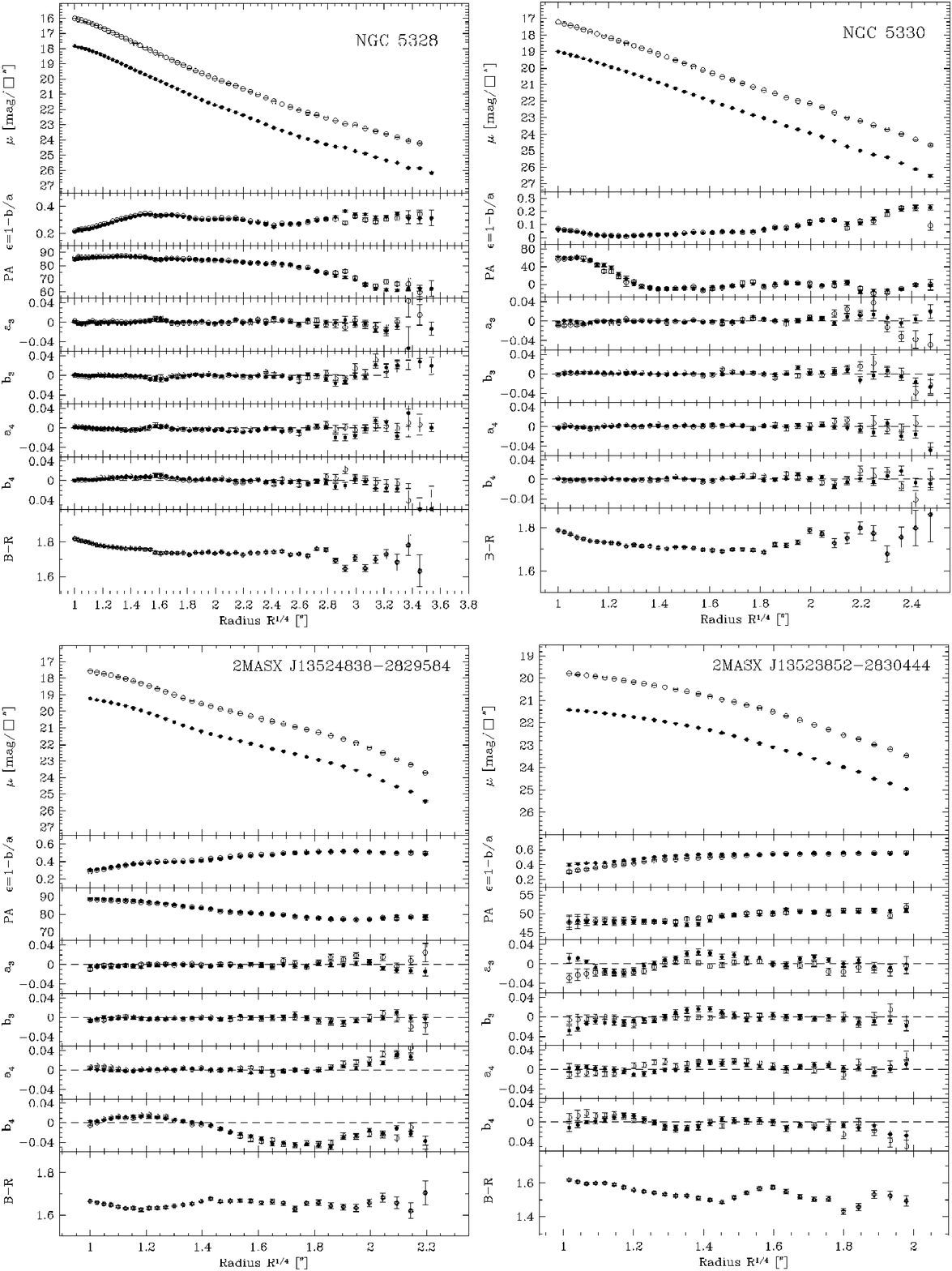}}
\caption{ B (full circles) and R (open circles) bands surface
photometry of group member galaxies. From top to bottom: surface
brightness ($\mu$), ellipticity ($\epsilon$), position angle profile
(P.A.), higher order coefficients of the Fourier expansion from the
interpolation of isophotes with ellipses (a$_3$, b$_3$, a$_4$, b$_4$)
and (B--R) colour profiles. In particular b$_4$ is the isophotal shape
parameter.
\label{fig4}}
\end{figure*}

\begin{figure*}
\resizebox{16cm}{!}{
\includegraphics[]{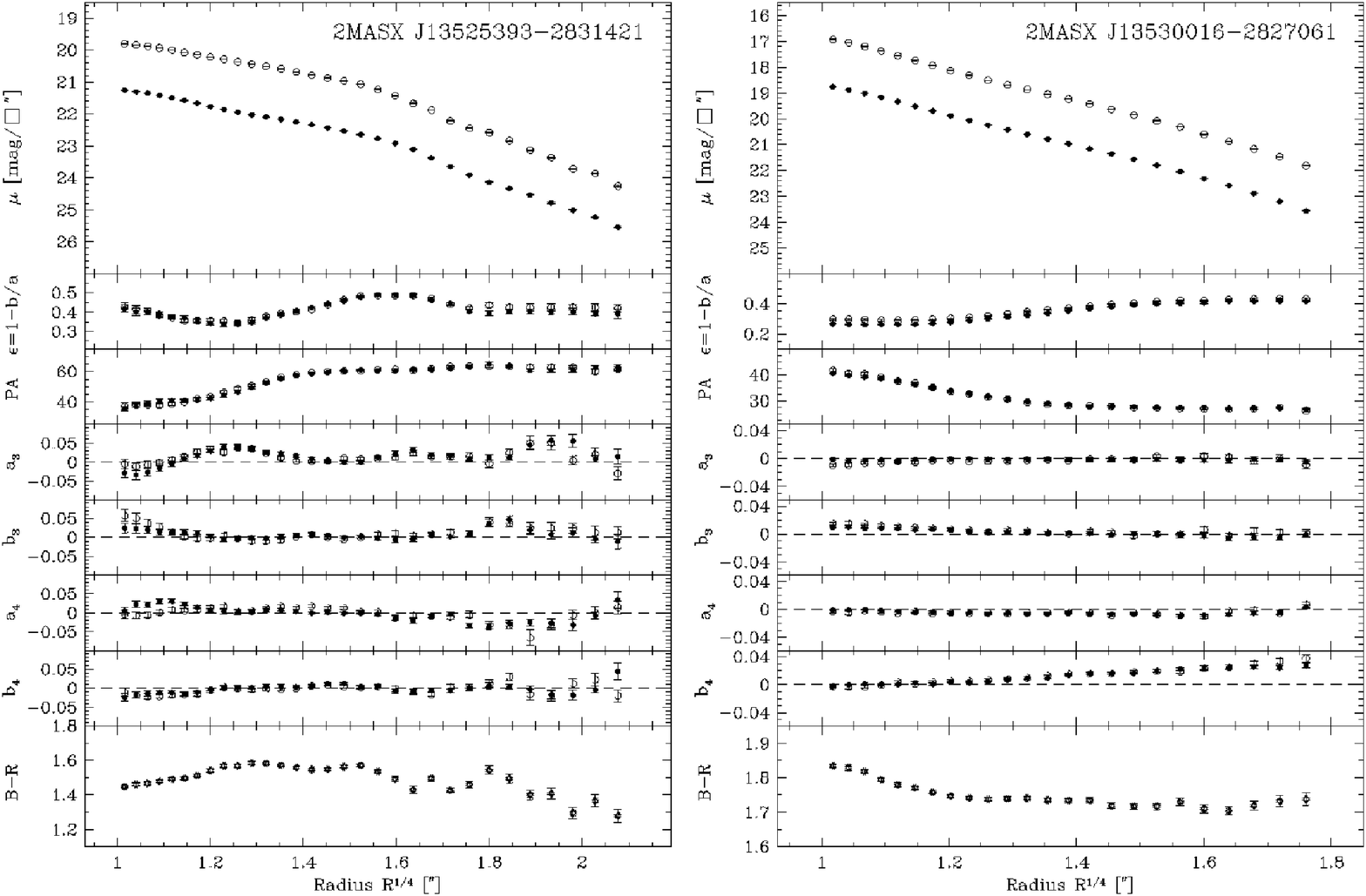}}
\caption{ As in Figure 4.
\label{fig5}}
\end{figure*}

\begin{figure}
\resizebox{7.5cm}{!}{
\includegraphics[]{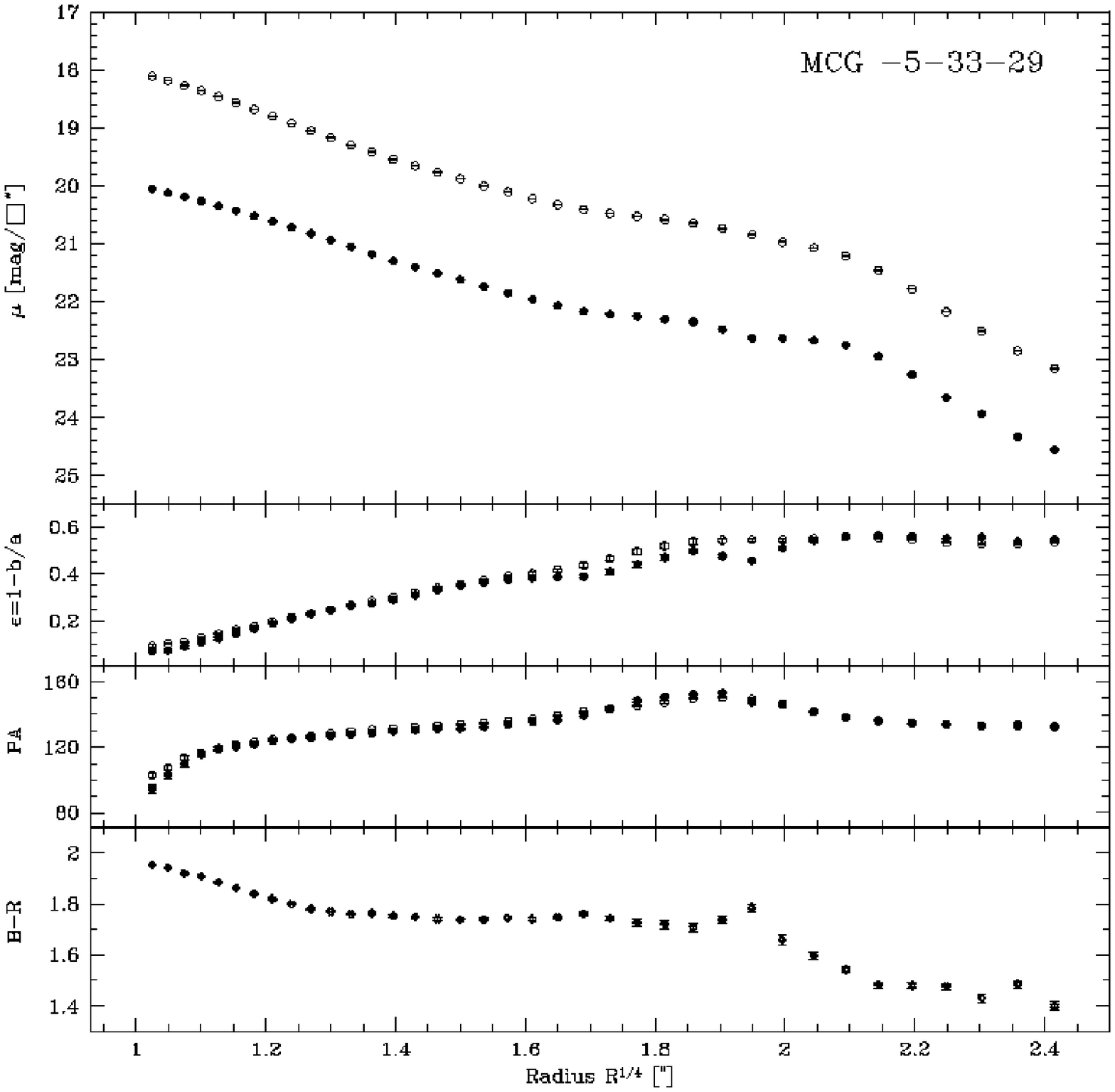}}
\caption{ As in Figure 4.
\label{fig6}}
\end{figure}

For each member galaxy the isophotal map, the B-band image and the
residual after a model subtraction to reveal fine structure, disc
components, spiral arms or other patterns are shown in
Figures~\ref{fig2} and \ref{fig3}. The surface photometry profiles
including surface brightness, ellipticity, position angle, the
coefficients of the Fourier expansion and the (B--R) colour profile
are displayed in Figures \ref{fig4}, \ref{fig5} and \ref{fig6}.

{\bf NGC 5328}~ The surface photometry (Figure~\ref{fig4}, top left)
shows a r$^{1/4}$ surface brightness profile over a range of 7 mag
suggesting that the galaxy is a ``bona fide'' elliptical.  The
ellipticity distribution increases in the range of $0.2 < \varepsilon
< 0.4$ up to a radius of 7\arcsec\ where it becomes flat. The $b_{4}$
profile is basically flat although a possible discy structure appears
in the residual image in Figure \ref{fig2} (top row, right panel). We
classified NGC~5328 as E3 in contrast to the classification as E1 by
\citet{rc3}. The colour profile is almost flat at a value of
(B--R)$\approx$ 1.7 mag.

{\bf NGC~5330}~ The surface brightness profile of this nearly round
elliptical was fitted by a Sersic power law with a high structural
parameter of $n=3.16$, with the isophotes showing no deviations from
pure ellipses (see Figure~\ref{fig4}, top right). The average
ellipticity is $\approx$ 0.1, while the position angle variation
outside the region affected by the seeing is less than
10$^\circ$. Like 2MASX J13524838-2829584, NGC~5330 is embedded in the
outer halo of NGC~5328 from which it is separated by 35~kpc in
projection. The luminosity and geometric profiles of NGC~5330 are
contaminated by the light of the nearby giant elliptical. The
contamination is probably the cause for the observed increase of the
(B--R) colour profile at radii $r > 16$\arcsec\ . The model subtracted
image shows a faint ripple-like residual at $ r \approx 10$\arcsec\
which does not seem to be an artifact since the isophotes in that
region -- as shown in Figure~\ref{fig4} -- are nearly perfect ellipses
and show virtually null variation of their position angle.

{\bf MCG~--5--33--29}~ The SBa spiral is located at about 124 kpc
projected distance from NGC~5328. The arms start from a bar at $r
\approx 10$\arcsec\ and appear tightly wound. The bar can be seen in
the flattening of the surface brightness profile at $r \approx
11$\arcsec. The (B--R) colour profile in the bar region is flat and
red ($\approx$1.75 mag) and drops to $\approx$1.4 mag in the position
of the arms and the ring. The model subtracted image
(Figure~\ref{fig3}) reveals the irregular patchy structure of the
spiral arms which both appear bifurcated.

{\bf 2MASX J13524838-2829584}~ The surface photometry of this small
galaxy is strongly contaminated by the vicinity of NGC~5328. The inner
parts are slightly discy while the outskirts are quite boxy ($b_4 =
-0.05$). The position angle variation of about 10$^{\circ}$ is only
marginally due to the light contamination caused by NGC~5328. The
colour profile is red and flat at (B--R)=1.65.

{\bf 2MASX J13523852-2830444}~ The morphology suggests a S0-type
galaxy seen almost edge-on. The presence of a disc is supported by the
increasing ellipticity $0.2 \leq \varepsilon \leq 0.6$ and the
constant position angle profiles with a variation lower than
5$^{\circ}$. The results from the surface photometry are affected by a
central major-axis dust lane, whose signature is evident in the
residual image (Figure \ref{fig2}) and in the positivity of the $b_4$
parameter.

{\bf 2MASX J13525393-2831421}~ We classify this galaxy as an SB0.
This classification is based upon the surface brightness profile
showing two components and the sudden isophotal twisting of about
20$^{\circ}$, indicating the presence of a bar component as suggested 
also in the isophotal image of Figure~\ref{fig3}. The (B--R) colour 
profile has an average value of 1.5 mag.

{\bf 2MASX J13530016-2827061}~ This galaxy is an edge--on S0 galaxy
with a weak disc component, merely distinguishable in the surface
brightness profile, but clearly visible in ellipticity, position angle
and the $b_4$ profile. The residual image in Figure~\ref{fig3}
shows the typical pattern of positive 4$ ^{th}$ Fourier--coefficients
indicative of strong discy isophotes. The disc is thickening with
increasing radius and is warped in its outer parts.

\begin{table*}
\caption{Photometric results}
\label{tab4}
\begin{tabular}{lccccccccccc} 
\hline
{object}    & {B$_T$}   & {M$_B$} & {R$_T$}   & {M$_R$}  & {r$_e$(B)} & 
{r$_e$(R)}
& {$\mu_e$(B)} & {$\mu_e$(R)} & {$\mu_0$(B)} & {$\mu_0$(R)} & {n} \\
 {(1)} & {(2)} & {(3)} & {(4)} & {(5)} & {(6)} & {(7)} & {(8)} & {(9)} & 
{(10)} & {(11)} & {(12)} \\
\hline
NGC~5328        & 12.31 & --21.85 & 10.52 & --23.64 & 53.16 & 58.31 & 24.00 & 
22.46 & 15.17 & 12.87 & 4.23 \\
NGC~5330        & 14.52 & --19.64 & 12.77 & --21.39 &  9.08 &  9.65 & 22.44 & 
20.88 & 15.92 & 13.72 & 3.16 \\
2MASX J13524838-2829584       & 15.48 & --18.68 & 13.84 & --20.32 &  5.94 &  5.90 & 21.80 & 
20.18 & 17.51 & 15.71 & 2.14 \\
2MASX J13523852-2830444       & 16.91 & --17.25 & 15.10 & --19.06 &  &  &  &  &  &  &  \\
~~bulge comp. & 18.02 & --16.14 & 17.73 & --16.43 & 
5.08 &  2.52 & 23.53 & 21.46 & 20.84 & 18.80 & 1.40 \\
2MASX J13525393-2831421       & 16.88 & --17.28 & 15.33 & --18.83 &  6.74 &  6.92 & 23.39 & 
21.64 & 21.05 & 19.28 & 1.24 \\
MCG~--5--33--29 & 14.69 & --19.47 & 13.05 & --21.11 &  &  &  &  &  &  &  \\
~~bulge comp. & 17.04 & --17.12 & 15.55 & --18.61 & 1.32 &  1.61 & 
20.13 & 18.94 & 18.31 & 17.83 & 1.00 \\
2MASX J13530016-2827061       & 15.60 & --18.56 & 13.87 & --20.29 &  &  &  &  &  &  &  \\
~~bulge comp. & 16.13 & --18.03 & 15.11 & --19.05 & 4.8  &  6.39 & 
21.68 & 21.16 & 19.27 & 20.05 & 1.27 \\

\hline
\end{tabular}

\medskip 
{$^1$}{~group member galaxy identification};
{$^2$}{~apparent total magnitude $m_B$}; 
{$^3$}{~absolute total magnitude $M_B$};
{$^4$)}{~apparent total magnitude $m_R$}; 
{$^5$}{~absolute total magnitude $M_R$};
{$^6$}{~effective radius r$_e$(B)[\arcsec]}; 
{$^7$}{~effective radius r$_e$(R)[\arcsec]};
{$^8$}{~surface brightness $\mu_e$(B) measured at the effective radius};
{$^9$}{~surface brightness $\mu_e$(R)}; 
{$^{10}$}{~central surface brightness $\mu_0$(B)};
{$^{11}$}{~central surface brightness $\mu_0$(R)}; 
{$^{12}$}{~Sersic index $n$}.
\end{table*}

\begin{table*}
\caption{Neighbour galaxies within 1 Mpc}
\label{tab5}
\begin{tabular}{lccllllc} \hline
{object} & {$\alpha$} & {$\delta$} & {type} & {m$_B$}  & {cz}               & 
{D$^1$} & {Notes$^2$} \\
              & (J2000.0) & (J2000.0) &            & [mag]     & 
[km~s$^{-1}$]& [kpc] & \\
\hline
2MASX J13521914--2839141   & 13$^h$52$^m$19.2$^s$ & --28$^{\circ}$39$'$14$''$ 
& S0          & 16.7  & 4669 $\pm$ 59  & 225  &  \\
2MASX J13514833--2823306   & 13$^h$51$^m$48.3$^s$ & --28$^{\circ}$23$'$31$''$ 
& S0          & 16.6  & 4970 $\pm$ 120 & 308  &  \\
2MASX J13530584--2847581   & 13$^h$53$^m$05.8$^s$ & --28$^{\circ}$47$'$58$''$
& S0          & 15.2  & 5148 & 308  &  \\
2MASX J13514400--2814526   & 13$^h$51$^m$44.0$^s$ & --28$^{\circ}$14$'$52$''$ 
& S0          & 16.1  & 5083 $\pm$ 76  & 418  &  \\
2MASX J13513401--2841297   & 13$^h$51$^m$34.0$^s$ & --28$^{\circ}$41$'$30$''$
& Sp          & 16.4  & 4407 & 420 &  \\
6dF J1354457--285159       & 13$^h$54$^m$45.6$^s$ & --28$^{\circ}$51$'$58$''$ 
& Sp          & 16.2  & 4645 $\pm$ 52  & 662  &  \\
2MASX J13531650--2900001   & 13$^h$53$^m$16.5$^s$ & --29$^{\circ}$00$'$00$''$ 
& Sp          & 16.7  & 4689 $\pm$ 45  & 697  &  \\
ESO 445 -- G 64            & 13$^h$52$^m$21.0$^s$ & --27$^{\circ}$53$'$39$''$ 
& SB(l)0/a    & 14.57 & 4907 $\pm$ 40  & 720  &  \\
ESO 445 -- G 51            & 13$^h$49$^m$20.9$^s$ & --28$^{\circ}$12$'$04$''$ 
& (R)SAB(r)ab & 14.56 & 5000 $\pm$ 10  & 959  & G \\
2MASX J13534359--2918241   & 13$^h$53$^m$43.6$^s$ & --29$^{\circ}$18$'$24$''$ 
& S           & 15.70 & 4998 $\pm$ 38  & 1009 &  \\
\hline
\end{tabular}

\medskip 
{$^1$}{~Projected distance from NGC~5328};
{$^2$}{~G -- Group member of LGG 357 according to Garcia (1993)}.
\end{table*}

\subsection{Spectroscopic results}

\begin{figure*}
\resizebox{18.0cm}{!}{
\includegraphics[]{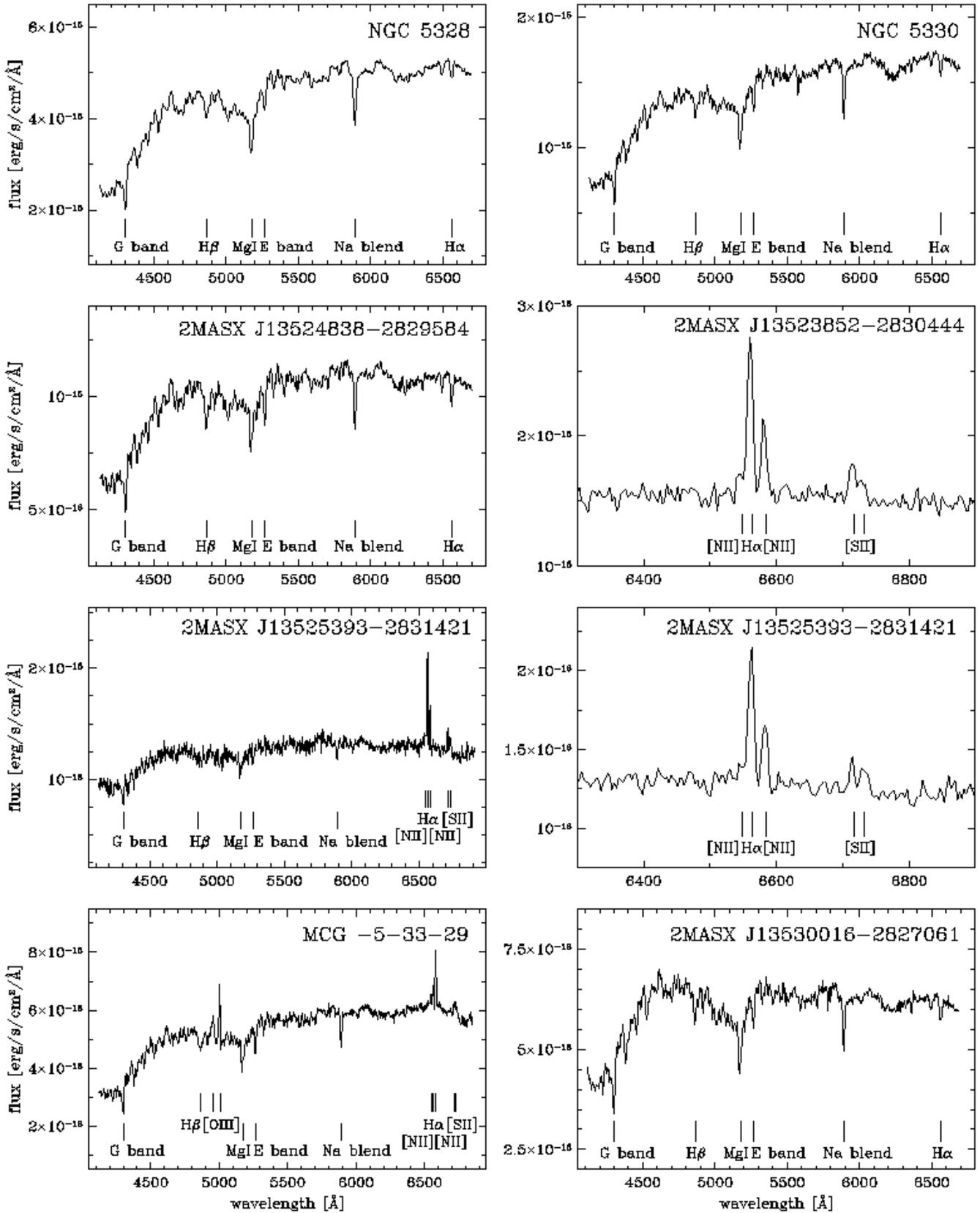}}
\caption{ Spectra of the NCG~5328 group member galaxies. 
\label{fig7}}
\end{figure*}

Table~3 collects the relevant information coming from the
spectroscopic study of the NGC~5328 group members.  Figure~\ref{fig7}
shows the spectra obtained from seven member galaxies. Both NGC~5328
and NGC~5330 show absorption--line spectra typical of elliptical
galaxies. The spectra of 2MASX J13524838-2829584 and 2MASX
J13530016-2827061 are consistent with their class and do not show
peculiarities.

Emission lines are shown by the remaining three galaxies. The spectrum
of the spiral MCG~--5--33--29 presents narrow but strong forbidden
lines of [O~III], [N~II] and [S~II].  H$\alpha$ is weak in emission
and H$\beta$ is weak in absorption.  Figure~\ref{fig7} shows the
emission features in 2MASX J13523852-2830444 in the region around
H$\alpha$ featuring narrow emission lines of H$\alpha$, [N~II] and
[S~II]. 2MASX J13525393-2831421 shows strong H$\alpha$ emission and
only weak [N~II] and [S~II] lines.

The average recession velocity of the group is 4937 (median 4785)
km~s$^{-1}$ with a velocity dispersion of $\sigma$ = 342 km s$^{-1}$,
while in the projected innermost part of the group, which includes
NGC~5328, NGC~5330 and 2MASX J13530016-2827061, the dispersion is
significantly lower (279 km s$^{-1}$).  The virial mass of the group
based on our redshift measurements and the projected separations
between each galaxy pair has been computed according to
\citet{hei85}. Although it is susceptible to systematic errors due to
the incompleteness of the sample, the probable non--equilibrium state
of the group or projection effects, it still represents a
straightforward and common mass estimator for galaxy groups. Inclusion
of galaxies lying wihtin 1~Mpc from NGC~5328 (see \ref{tab5}) yielded
$M_{VT} = 1.6\times 10^{13} ~M_\odot$, which is a typical value for
poor galaxy groups.

\subsection {Properties of emission lines and absorption line-strength indices}

\begin{figure}
\resizebox{8cm}{!}{
\includegraphics[]{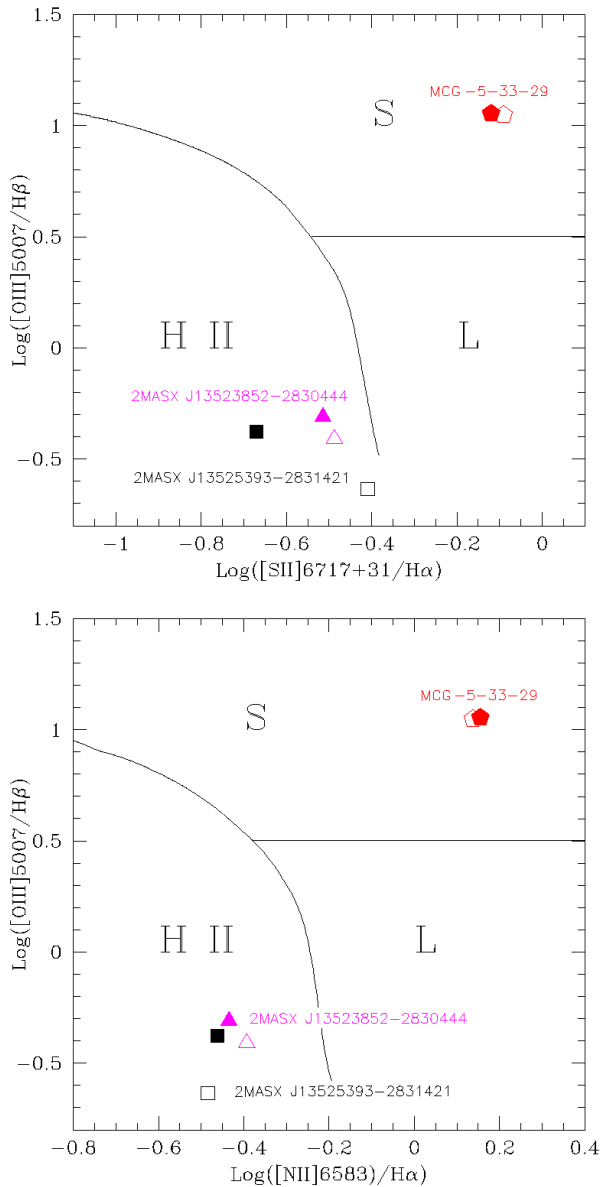}}
\caption{Diagnostic diagrams for emission--line galaxies: logarithmic
line ratios of [S~II] $\lambda 6617+31$/H$\alpha$ vs. [O~III] $\lambda
5007$/H$\beta$ (top panel) and [N~II] $\lambda$6583/H$\alpha$
vs. [O~III] $\lambda 5007$/H$\beta$ (bottom panel). The emission--line
flux was extracted within an aperture with radius R$_e/8$ (open
symbols) and R$_e/4$ (full symbols).  The bold line separates H~II
region like objects from AGNs.
\label{fig8}}
\end{figure}

The properties of the detected emission lines were used to
characterize and classify the group members. Emission line EWs have
been measured by modeling the underlying stellar component and
subtracting it to each galaxy spectrum (see also next paragraphs). The
derived emission--line ratios are plotted in the diagnostic diagrams
proposed by \citet{vei87} shown in Figure~\ref{fig8}. According to
these diagrams, the spiral MCG~--5--33--29 has emission properties
typical of an AGN while both the S0s 2MASX J13523852-2830444 and 2MASX
J13525393-2831421 present emissions features characteristic of HII
regions, suggesting the presence of ongoing or recent star formation
in their centers.

\begin{figure*}
\resizebox{18cm}{!}{
\includegraphics[]{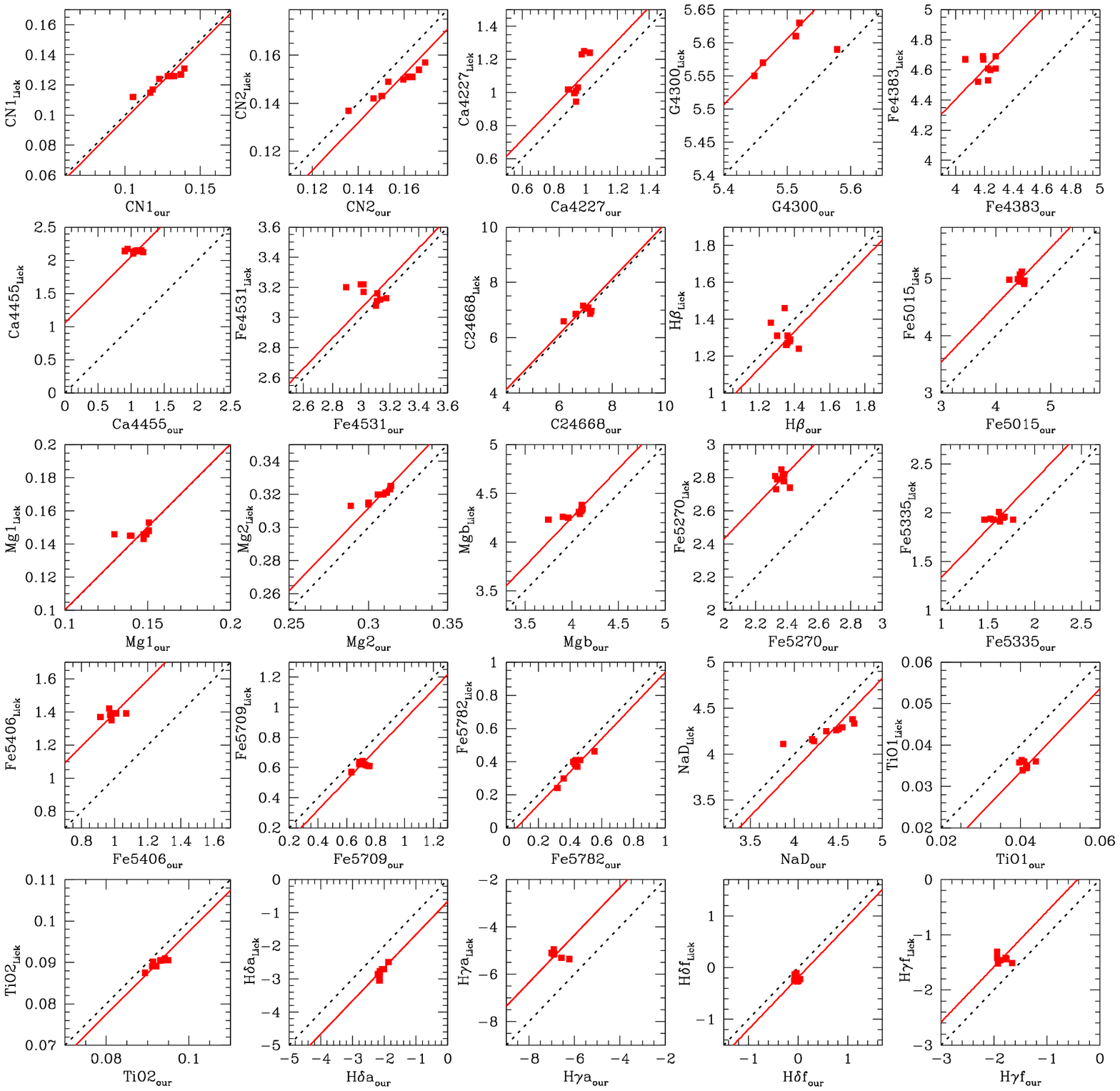}}
\caption{Comparison of passband measurements from our spectra of NGC 5328 and 
Lick calibrated apertures for the same galaxy in \citet{ramp05} . The dotted
line is the one to one relation while the solid line is the {\it robust
straight-line fit} to the filled squares (see text and \citet{ramp05}).
\label{fig9}}
\end{figure*}

Recently \citet{ramp05} presented a study of line-strength indices
calibrated in the Lick-IDS system \citep{WO97} for a sample of
early-type galaxies which also includes NGC~5328. Using the spectrum
of NGC~5238 as reference, the same calibration procedure was applied
to remaining galaxy spectra of the group as described in
\citep{ramp05}. The full procedure includes the correction for
possible hydrogen emission, correction for velocity dispersion and the
transformation to the Lick-IDS system.  
We are aware that the standard calibration to the Lick system lies
on the observation and the comparison with a set of Lick-IDS standard stars.
For this transformation Lick-IDS standard stars have been observed 
but yielded unsatisfactory calibration results.  We then
approached the transformation to the Lick-system by comparing our
raw indices of NGC~5328, measured for 7 different apertures, with the
corresponding corrected line-strength indices for the same galaxy obtained
by \citet{ramp05}. The comparison is shown in Figure~\ref{fig9}.
In the literature \citep[see also][]{Puzia2002} the transformation to the Lick
system is computed by fitting the points with a linear relation of the 
kind EW$_{Lick}$=$\alpha$~EW$_{raw}$ + $\beta$. For most of the indices the
slopes of the transformation are very close to 1  and only a zero point
offsets are required. Given the limited number of points, the small variation
of each index, and the "non standard" procedure, we calculated the correction to
the Lick system forcing $\alpha$=1 in the fit and deriving only the offsets $\beta$.
In Figure~\ref{fig9} the  dotted line is the one to one relation, while
the solid line marks the shift to apply for the transformation into the Lick
system. Within the above set of working hypotheses, we derive the final corrected
line strength indices for two radii, r$_{e}$/8 and r$_{e}$/4, for the four group
galaxy members.

\begin{figure*}
\resizebox{18cm}{!}{
\includegraphics[]{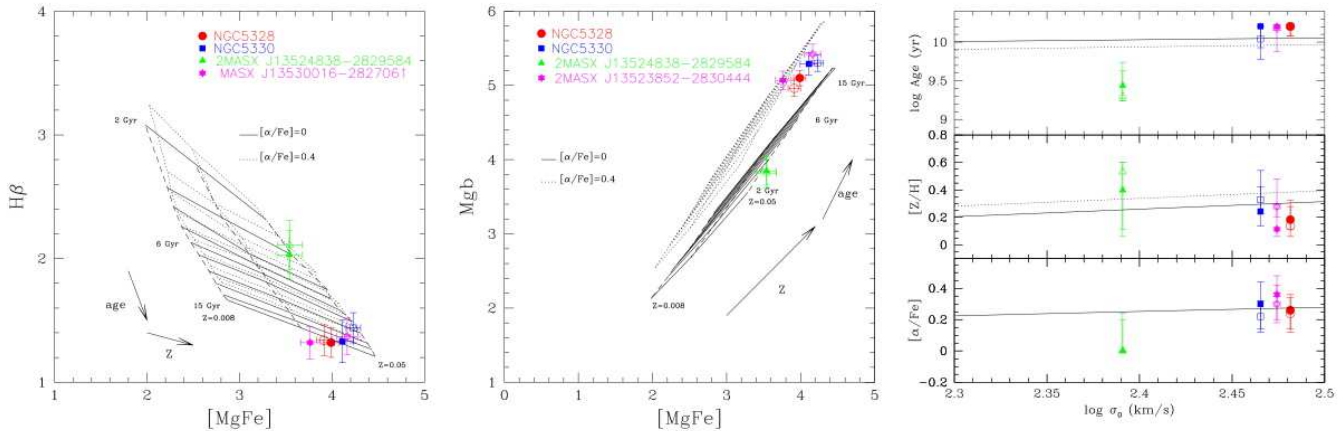}}
\caption{{\bf left panel}: H$\beta$ vs. [MgFe] plane where we plot the
measured indices for two apertures (R$_e/8$, solid symbols, and
R$_e/4$, open symbols) for those galaxies with no emission lines,
together with SSP models with $0.008<Z<0.05$, 2 Gyr $<t<$ 15 Gyr and
$[\alpha/Fe]=$0 and 0.4. Solid ``horizontal'' lines indicate
solar-scaled models of constant age, while ``vertical'' dashed lines
are models of constant metallicity. Dotted lines represent
$\alpha$-enhanced SSPs. ({\bf mid panel}): Mgb vs. MgFe plane. Lines
of constant age and metallicity are degenerate and run almost
parallel. The effect of $\alpha$-enhancement is instead well separated
in this plane.  ({\bf right panel}) From top to bottom we plot the
derived ages, metallicities and $[\alpha/Fe]$ ratios for the NGC~5328
group galaxy members which do not show strong emission lines, namely
NGC~5328, NGC~5330, 2MASX J13530016-2827061 and 2MASX
J13524838-2829584. Different symbols refer to different objects and
apertures with the same meaning as in the previous panels. The dotted
and solid lines represent the best fit for the low and high density
environments obtained as function of the velocity dispersion by
\citet{thom05}. In the $[\alpha/Fe]$ vs. $log~\sigma$ plot the 
two environments share the same best fit relation.
\label{fig10}}
\end{figure*}

The correction of hydrogen absorption lines for emission infilling
deserves a special comment.  In particular, the measure of the
H$\beta$ index of the underlying stellar population could be
contaminated by a significant infilling due to the H$\beta$ emission
component. We used the same template galaxy (NGC~1426) adopted in
\citet{ramp05} to derive the [OIII]($\lambda$5007\AA) emission EW for
the member galaxies and applied the correction
EW(H$\beta_{em}$)/EW([OIII])=0.7 proposed by \citet{gonz93}.
Of the four galaxies shown in Figure~\ref{fig10} the only 
correction worth to mention is that applied to NGC~5330. The H$\beta$
indices for r$_{e}$/8 and r$_{e}$/4 move from 1.20 to 1.33 and
1.40 to 1.44 respectively after correction for emission infilling.

In order to characterize the underlying stellar population of the
galaxies we have compared the fully corrected line-strength indices
with Simple Stellar Population (SSPs) for a wide range of
metallicities ($0.0004 <Z<0.05)$, ages ($1 Gyr<t<15 Gyr$) and
$\alpha$-element enhancements ($0<[\alpha/Fe]<0.4$).  For a full
description of the models we refer to \citet{Anni05a} and
\citet{Anni05b}.  The solar-scaled composition SSPs have been derived
according to the procedure described in \citet{bre96} and on the index
passbands definition of \citet{WFGB94} and \citet{WO97}.  The
solar-scaled composition SSPs have been derived according to the
procedure described in \citet{bre96}, while $\alpha$-enhanced models
are based on the \citet{TB95}, response functions and on new responses
derived by model atmospheres and synthetic stellar spectra computed
with the code ATLAS9 \citep{kur93}.  In the left and mid panels of
Figure~\ref{fig10} we show respectively the classic H$\beta$ vs. [MgFe]
plane and the Mgb vs. [MgFe] plane where we plot the measured galaxy
indices for the apertures r$_{e}$/8 (full symbols) and r$_{e}$/4 (open
symbols) together with model SSPs for different ages ($2 Gyr <t< 15
Gyr$), metallicites ($0.008<Z<0.05$) and enhancements
([$\alpha$/Fe]=0, solid-dashed lines; [$\alpha$/Fe]=0.4, dotted
lines).  The Seyfert MCG-5-33-29 and the S0s 2MASX J13523852-2830444
and 2MASX J13525393-2831421 are not included in the analysis since
these galaxies have strong emission lines and the quality of our
spectra does not allow to recover a reliable value of the H$\beta$
absorption index.

The H$\beta$ vs. [MgFe] plane, where models of constant age (solid
lines) and constant metallicity (dashed lines) run almost
orthogonally, is usually considered a powerful tool to disentangle age
and total metallicity (see e.g. \citet{W94}, \citet{WO97}).  In this
plane the $\alpha$-elements enhancement has only a small effect on the
SSPs: the [$\alpha/Fe$]=0.4 models, represented by the dotted line,
preserve the same [MgFe] values and are only slightly shifted toward
stronger $H\beta$.  On the other hand, the effect of non-solar
partitions is well visible in the Mgb vs. [MgFe] plane. For a fixed
element composition, lines of constant age and metallicity are
basically degenerate and run almost parallel, while changes in the
[$\alpha$/Fe] ratio produce a considerable shift on the models.

From the combined analysis of the two planes the following
considerations can be done: NGC~5328, NGC~5330 and 2MASX
J13530016-2827061 have quite old stellar populations, super-solar
metallicities and super-solar [$\alpha$/Fe] ratios. 2MASX
J13524838-2829584 presents a super-solar metallicity as well, but no
$\alpha$-enhancement. The strong $H\beta$ value however suggests the
occurrence of a recent burst of star formation.  We are pretty well
confident about this result because the galaxy is not affected by
emission and the H$\beta$ index is therefore reliable.  For what
concerns the large error on the measured velocity dispersion, it
mainly affects the velocity dispersion correction for the [MgFe] and
Mgb indices, but has almost no effect on the H$\beta$.  This
translates into a larger uncertainty in the derived metallicity than
in the derived age, which is always $<$ 4 Gyr.

In order to derive SSP parameters (age, Z and [$\alpha$/Fe]) we have
devised a simple but robust algorithm based on the Mgb, $<Fe>$ and
H$\beta$ indices (\citet{Anni05b}).  For each galaxy the best-fit age,
Z and [$\alpha$/Fe] are derived by searching in a finely spaced grid
of points in the (H$\beta$, Mgb and $<Fe>$) space the SSP model that
matches the observed indices.  In Fig.~\ref{fig10} (right panel) the
derived SSP parameters with the associated errors are plotted as a
function of central velocity dispersion for the galaxies NGC~5328,
NGC~5330, 2MASX~J13530016-2827061 and 2MASX~J13524838-2829584.
The dotted and solid lines  represent the best fit for the low and high
density environments obtained as function of the velocity dispersion by
\citet{thom05}. In the $[\alpha/Fe]$ vs. $log~\sigma$ plot the 
two environments share the same best fit relation. Both the trend and
the dispersion of our values are similar to those found in the set of object
studied in low and high density environments by \citet{thom05}. 


\section{Discussion and conclusions}

We analyzed photometric and spectroscopic properties of seven
galaxies members of the NGC~5328 group of galaxies. We analyzed the
galaxy structure, studied the emission line properties and attempted
to infer the star formation history of the galaxies in the group from
the comparison of the measured line-strength indices with SSP models.

\begin{figure}
\resizebox{8cm}{!}{
\includegraphics[]{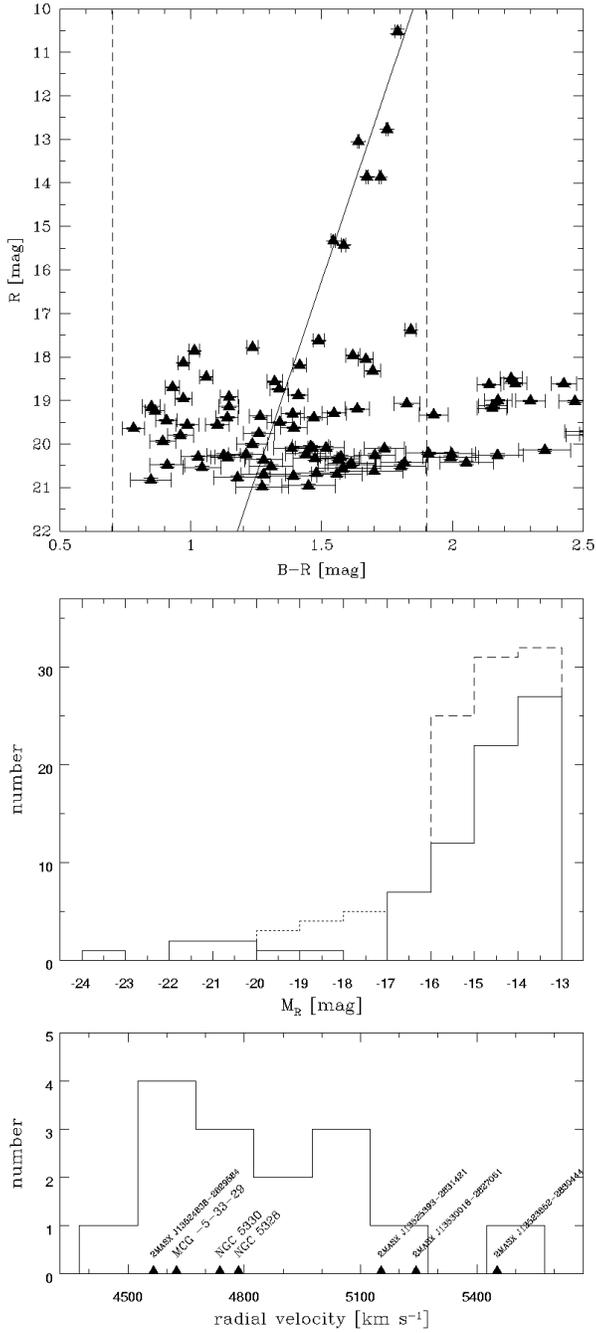}}
\caption{(upper panel) Colour--magnitude relation for the group
including bright galaxies at the top and the candidate dwarf galaxies
in the field of the group. Dwarf galaxy candidates are enclosed within
the dashed lines, the solid line represents the colour--magnitude
relation for the Coma cluster (see text). The list of the objects is
given in Appendix A.  (mid panel) Absolute R magnitude distribution
within the group including all extended objects (long dashed line),
dwarf member candidates after applying the colour restriction (solid
line) and neighbour galaxies within 1 Mpc (dotted line). (bottom
panel) Redshift distribution of known group members, including those
detected in this paper, and of neighbour galaxies within 1 Mpc.  
\label{fig11}}
\end{figure}

The NGC~5328 group represents a relatively compact structure of
galaxies lying in the outskirts of a galaxy cluster: 
It consists of seven galaxies aligned in a chain spanning about 200~kpc.
The group has the same redshift of Abell~3574 which has a projected
distance of 2.4~Mpc. It appears thus particularly suited to investigate
processes occuring in galaxy groups which are likely to be accreted by
galaxy clusters. A set of 10 neighbour
galaxies with known redshift within 1 Mpc (see Table~\ref{tab5}) has
been extracted from {\tt NED}. These objects do not show a symmetrical
projected distribution around the group, their redshift distribution
is plotted (together with the confirmed group members) in
Figure~\ref{fig11} (bottom panel). We notice a sort of morphology
segregation in the NGC~5328 group: two ellipticals are forming the
center of the group (in projection and in redshift space) while the
lenticulars are located from small to medium distances, the spiral
MCG~--5--33--29 has the largest projected separation from NCG~5328. No
elliptical galaxies are found in the outskirts of the group: the
neighbour galaxies in Table~\ref{tab5} are faint spiral or lenticular
galaxies.  In the field covered by our images we investigated the
presence of a dwarf galaxy population. Figure~\ref{fig11} displays the
R-band magnitude distribution for the sample of candidate dwarfs and
the colour--magnitude relation of the Coma cluster \citep{sec97}
superposed to our data. Since this relation perfectly fits our data,
the same colour restriction for probable dwarf galaxy group members of
0.7 $\leq$ (B--R) $\leq$ 1.9, representing generous limits to the
early-type sequence, was applied to our sample. For the Coma cluster,
these limits were found to remove the vast majority of background
objects. The resulting sample of likely dwarf galaxy group members
consisting of 68 objects with a mean colour of (B--R) = 1.32 $\pm$
0.28 magnitudes is given in the Appendix, Tables~A1 and A2.

\begin{figure}
\resizebox{8cm}{!}{
\includegraphics[]{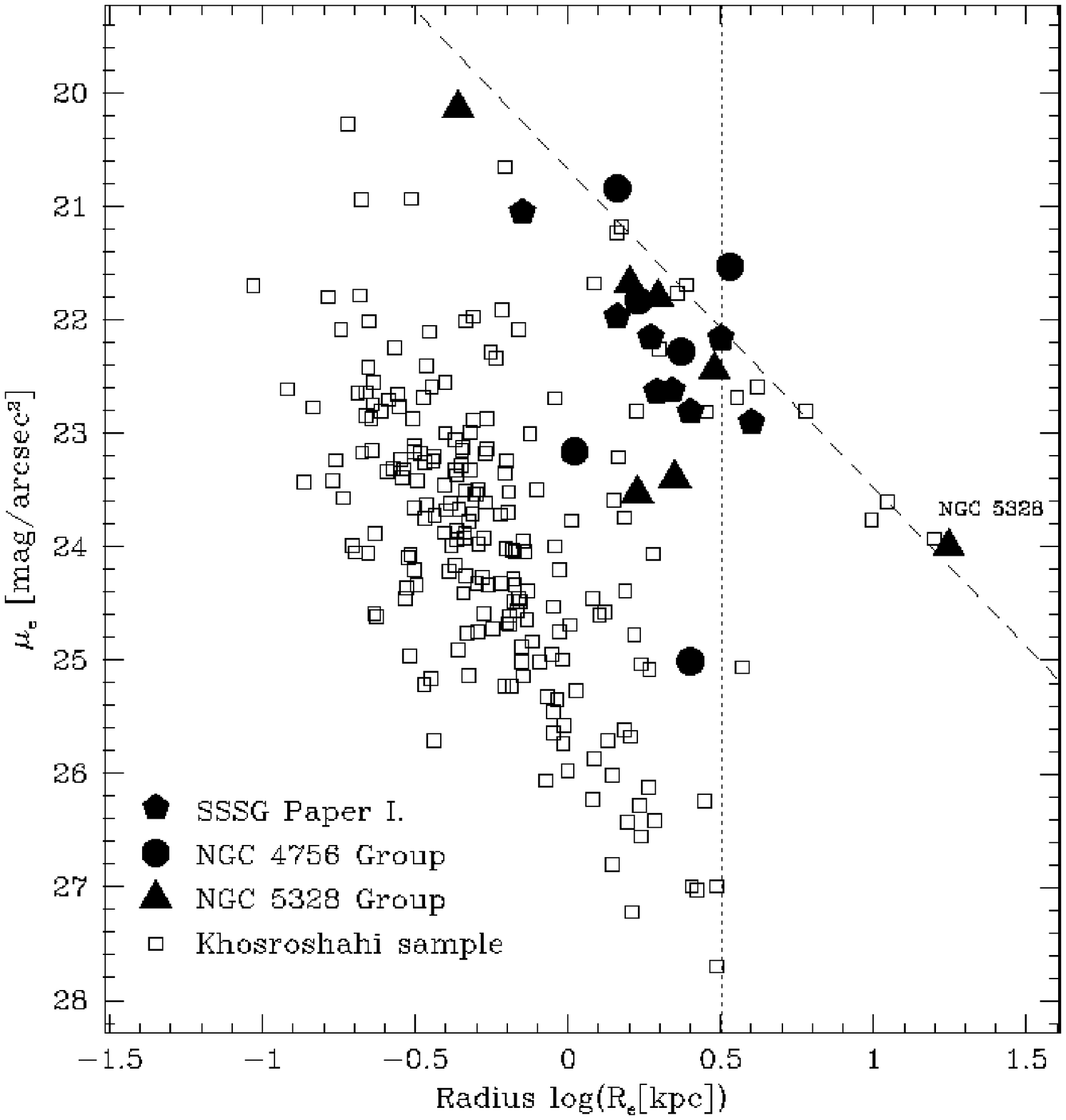}}
\caption{Effective radius (in kiloparsec) vs. the effective B surface
brightness. The dia\-gonal long--dashed line marks the HK87 relation
followed by elliptical galaxies and early--type bulges. The vertical
dotted line at log R$_e$ = 0.5, i.e., 3 kpc (H$_0$ = 70 km
s$^{-1}$~Mpc$^{-1}$) separates bright from ordinary ellipticals
according to \citet{cap92}. The plot shows the present data
(triangles) overplotted to measurements obtained in paper~I and
paper~II (pentagons).  Open squares are from the groups studied by
\citet{kho04} who used a similar procedure selecting dwarf galaxy
candidates by means of the R-(B--R) colour--magnitude relation.
\label{fig12}}
\end{figure}

The presence of a large number of dwarf galaxies in poor groups is
also suggested by \citet[][and references therein]{kho04} looking for
differences in the galaxy population of X-ray dim and X-ray bright
nearby groups. The photometric characteristic of the groups can be
easily shown using the Hamabe--Kormendy relation, a photometric
projection of the fundamental plane connecting the effective radius
R$_e$ with the surface brightness at this radius $\mu_e$. This
relation can be used to divide early--type galaxies in ordinary and
bright families \citep{ham87,cap92}. Bright elliptical galaxies follow
the relatively tight relation of $\mu_e$ = 2.94~log R$_e$ + 20.75
while the ordinary objects are spread over the region below this
relation with a borderline at R$_e$ = 3~kpc. This ordinary class is
believed to participate in the formation of the bright ellipticals
following the HK relation via successive merging episodes, although
the recent simulations of \citet{evs04} suggest that only progenitors
of higher surface brightness from the ordinary object class can
produce an outcome lying on the HK relation. Figure \ref{fig12} shows
the Hamabe--Kormendy relation for galaxies of the NGC~5328 group
together with data obtained in paper I and II for different
small--scale systems of galaxies compared with the galaxy group sample
studied by \citet{kho04}. Most of the early-type galaxies in groups
are ordinary objects, while NGC~5328 is located in the region of
bright galaxies, suggesting an advanced stage in its evolution.

Signatures of past and/or recent interactions between the group
members have been investigated looking for fine structures in the
galaxies and analyzing their line-strength indices. Only minor fine
structures have been detected: a faint ripple is found in NGC~5330,
however, no recent star formation episodes are suggested for this
galaxy by the H$\beta$ vs. MgFe line-strength indices.  The extremely
boxy isophotes of 2MASX J13524838-2829584 are believed to be connected
to a merging event \citep{sch96}.  Line strength indices suggest that
this object may have experienced a recent star formation episode. A
warped disc component emerges from the model subtracted image of 2MASX
J13530016-2827061. Since \citet{nel95} conclude that warps cannot be
primordial structures but must have either a recent or a continuous
excitation mechanism, we suggest an ongoing interaction of this galaxy
with nearby group members.

Star formation and nuclear activity seem present only in the projected
outskirts of the group. Spectral signatures of star formation are
present in the galaxies 2MASX J13523852-2830444 and 2MASX
J13525393-2831421.  Nuclear activity is present in MCG~--5--33--29:
the strong forbidden lines of [O~III] and [N~II] ratios indicate a
type 2 Seyfert activity.

Recent results \citep{Proctor04,Mendes05} suggest that galaxies 
in Hickson compact group (HCG) are generally old, a result that contradicts the
expectations from various scenarios for compact group formation. 
In particular, \citet{Diaferio94} and \citet{ Governato96} suggest that
a large fraction of ellipticals in compact groups should have a merging
origin and then should consequently appear to be young which is at odd
with above observations.  
NGC~5328 has not the characteristics of a compact group but
its has some similarity with them. In particular, the present study shows that
this group is composed of a large fraction of early-type members, as HCG
\citep{Hickson88}.  Futhermore, the dominant ``bona fide'' elliptical
NGC~5328 and the bright companion NGC~5330 have a quite old stellar population
and less luminous nearby early-type companions show both, signatures of recent
and ongoing star formation. We may depict NGC~5328 as an {\it evolving group}
with a plethora of possibly associated dwarf galaxies, which could be among the
drivers of the main galaxy evolution.

Most of the NGC~5328 group features are shared with poor groups
with X-ray diffuse emission \citep[see e.g.][]{mul00}. RASS reveals
that NGC~5328 is a relatively strong X--ray emitter \citep{beu99}: new
X--ray observations would clarify the nature and the characteristics
of the emission (e.g. the presence of a ``group'' component in the hot
diffuse medium \citep{mul00} yielding valuable clues about the
evolutionary phase of this group.

\section*{Acknowledgments}
RG and WWZ acknowledge the support of the Austrian Science Fund
(project P14783). RG, RR and WWZ acknowledge the support of the
Austrian and Italian Foreign Offices in the framework science and
technology bilateral collaboration (project number 25/2004) This
research has made use of the NASA/IPAC Extragalactic Database ({\tt
NED}) which is operated by the Jet Propulsion Laboratory, California
Institute of Technology, under contract to the National Aeronautics
and Space Administration. The Digitized Sky Survey (DSS) was produced
at the Space Telescope Science Institute under U.S.  Government grant
NAG W-2166. The images of these surveys are based on photographic data
obtained using the Oschin Schmidt Telescope on Palomar Mountain and
the UK Schmidt Telescope. The plates were processed into the present
compressed digital form with the permission of these institutions.

\appendix
\section{Tables of faint group member candidates}
\begin{table*}
\caption{Faint  group member candidates}
\label{tab6}

\begin{tabular}{rccccc} 
\hline
 {object$^1$}   & {$\alpha$} & {$\delta$} & {m$_R$}   & {M$_R$}   & {(B--R)} \\
                & (J2000.0) & (J2000.0)& [mag]     & [mag]        & [mag]   \\
\hline
767 & 13 53 14.045 & -28 25 49.178 & 17.380 $\pm$ 0.008 & -16.78 & 1.841 $\pm$ 
0.022 \\
343 & 13 53 02.685 & -28 30 16.420 & 17.617 $\pm$ 0.012 & -16.55 & 1.488 $\pm$ 
0.022 \\
1105 & 13 53 07.950 & -28 27 53.585 & 17.785 $\pm$ 0.010 & -16.38 & 1.236 
$\pm$ 0.021 \\
462 & 13 52 47.719 & -28 30 04.422 & 17.856 $\pm$ 0.010 & -16.31 & 1.014 $\pm$ 
0.020 \\
1203 & 13 53 14.474 & -28 28 48.712 & 17.966 $\pm$ 0.011 & -16.20 & 1.619 
$\pm$ 0.026 \\
680 & 13 52 49.206 & -28 28 46.751 & 18.055 $\pm$ 0.012 & -16.11 & 1.669 $\pm$ 
0.029 \\
749 & 13 53 11.407 & -28 25 05.067 & 18.137 $\pm$ 0.011 & -16.03 & 0.971 $\pm$ 
0.022 \\
159 & 13 52 36.845 & -28 32 28.946 & 18.189 $\pm$ 0.012 & -15.97 & 1.416 $\pm$ 
0.027 \\
1007 & 13 52 56.378 & -28 27 02.921 & 18.318 $\pm$ 0.013 & -15.84 & 1.695 
$\pm$ 0.032 \\
86 & 13 52 39.197 & -28 32 59.800 & 18.439 $\pm$ 0.013 & -15.72 & 2.97 $\pm$ 
0.057 \\
634 & 13 52 58.040 & -28 29 03.726 & 18.461 $\pm$ 0.014 & -15.70 & 1.059 $\pm$ 
0.027 \\
492 & 13 52 57.540 & -28 29 51.576 & 18.495 $\pm$ 0.014 & -15.67 & 2.224 $\pm$ 
0.042 \\
712 & 13 53 13.512 & -28 28 19.031 & 18.507 $\pm$ 0.014 & -15.66 & 2.643 $\pm$ 
0.051 \\
676 & 13 53 13.445 & -28 28 46.790 & 18.569 $\pm$ 0.014 & -15.59 & 1.32 $\pm$ 
0.030 \\
784 & 13 53 05.695 & -28 25 11.204 & 18.588 $\pm$ 0.014 & -15.57 & 2.669 $\pm$ 
0.054 \\
724 & 13 53 02.792 & -28 27 19.298 & 18.604 $\pm$ 0.014 & -15.56 & 2.24 $\pm$ 
0.045 \\
667 & 13 53 13.213 & -28 28 44.546 & 18.620 $\pm$ 0.015 & -15.54 & 2.425 $\pm$ 
0.048 \\
1147 & 13 52 55.937 & -28 26 56.279 & 18.636 $\pm$ 0.016 & -15.53 & 2.14 $\pm$ 
0.046 \\
717 & 13 53 10.249 & -28 24 49.202 & 18.698 $\pm$ 0.015 & -15.46 & 0.93 $\pm$ 
0.028 \\
564 & 13 52 57.758 & -28 28 52.981 & 18.722 $\pm$ 0.016 & -15.44 & 3.067 $\pm$ 
0.071 \\
1022 & 13 53 08.451 & -28 26 49.969 & 18.735 $\pm$ 0.016 & -15.43 & 1.337 
$\pm$ 0.034 \\
274 & 13 52 38.762 & -28 31 39.396 & 18.876 $\pm$ 0.017 & -15.29 & 1.411 $\pm$ 
0.037 \\
43 & 13 52 38.772 & -28 33 34.629 & 18.916 $\pm$ 0.017 & -15.25 & 1.146 $\pm$ 
0.034 \\
373 & 13 53 01.384 & -28 30 45.466 & 18.936 $\pm$ 0.018 & -15.23 & 2.736 $\pm$ 
0.066 \\
10 & 13 52 52.445 & -28 33 56.620 & 18.960 $\pm$ 0.017 & -15.20 & 0.971 $\pm$ 
0.033 \\
93 & 13 52 50.584 & -28 33 07.252 & 18.998 $\pm$ 0.018 & -15.16 & 2.174 $\pm$ 
0.053 \\
831 & 13 53 05.104 & -28 25 43.464 & 19.011 $\pm$ 0.018 & -15.15 & 2.298 $\pm$ 
0.056 \\
1148 & 13 52 45.610 & -28 27 01.086 & 19.020 $\pm$ 0.024 & -15.14 & 2.468 
$\pm$ 0.063 \\
890 & 13 53 00.848 & -28 26 01.148 & 19.073 $\pm$ 0.021 & -15.09 & 1.825 $\pm$ 
0.050 \\
1140 & 13 52 47.173 & -28 28 21.117 & 19.121 $\pm$ 0.021 & -15.04 & 2.155 
$\pm$ 0.057 \\
533 & 13 53 14.715 & -28 28 36.039 & 19.140 $\pm$ 0.019 & -15.02 & 0.849 $\pm$ 
0.035 \\
584 & 13 53 07.866 & -28 28 56.519 & 19.141 $\pm$ 0.020 & -15.02 & 1.146 $\pm$ 
0.039 \\
835 & 13 53 04.508 & -28 25 41.366 & 19.179 $\pm$ 0.019 & -14.98 & 2.152 $\pm$ 
0.057 \\
523 & 13 52 57.716 & -28 28 32.220 & 19.198 $\pm$ 0.020 & -14.96 & 1.636 $\pm$ 
0.047 \\
416 & 13 52 42.450 & -28 30 23.788 & 19.206 $\pm$ 0.022 & -14.96 & 2.702 $\pm$ 
0.076 \\
903 & 13 53 03.593 & -28 25 59.642 & 19.242 $\pm$ 0.020 & -14.92 & 0.863 $\pm$ 
0.036 \\
833 & 13 53 13.943 & -28 27 44.427 & 19.291 $\pm$ 0.020 & -14.87 & 1.547 $\pm$ 
0.047 \\
660 & 13 52 46.516 & -28 29 35.698 & 19.306 $\pm$ 0.021 & -14.86 & 1.389 $\pm$ 
0.045 \\
710 & 13 53 04.732 & -28 24 37.850 & 19.334 $\pm$ 0.021 & -14.83 & 1.928 $\pm$ 
0.056 \\
765 & 13 53 16.571 & -28 24 54.930 & 19.356 $\pm$ 0.021 & -14.81 & 2.958 $\pm$ 
0.089 \\
886 & 13 53 09.681 & -28 26 08.411 & 19.362 $\pm$ 0.021 & -14.80 & 1.265 $\pm$ 
0.043 \\
893 & 13 53 10.600 & -28 25 37.048 & 19.392 $\pm$ 0.022 & -14.77 & 1.139 $\pm$ 
0.043 \\
144 & 13 52 43.840 & -28 32 41.331 & 19.395 $\pm$ 0.027 & -14.77 & 1.47 $\pm$ 
0.052 \\
339 & 13 52 56.643 & -28 31 06.975 & 19.419 $\pm$ 0.022 & -14.74 & 2.713 $\pm$ 
0.082 \\
443 & 13 52 56.447 & -28 30 18.392 & 19.459 $\pm$ 0.022 & -14.70 & 0.907 $\pm$ 
0.040 \\
228 & 13 52 48.345 & -28 31 59.775 & 19.504 $\pm$ 0.023 & -14.66 & 1.34 $\pm$ 
0.049 \\
735 & 13 53 00.874 & -28 24 52.194 & 19.557 $\pm$ 0.024 & -14.61 & 0.986 $\pm$ 
0.044 \\
594 & 13 52 49.465 & -28 29 07.259 & 19.558 $\pm$ 0.025 & -14.60 & 1.102 $\pm$ 
0.046 \\
828 & 13 53 07.375 & -28 25 31.949 & 19.625 $\pm$ 0.024 & -14.54 & 1.394 $\pm$ 
0.052 \\
1065 & 13 52 44.363 & -28 27 27.749 & 19.641 $\pm$ 0.024 & -14.52 & 0.781 
$\pm$ 0.042 \\
933 & 13 53 10.738 & -28 26 30.416 & 19.711 $\pm$ 0.026 & -14.45 & 2.514 $\pm$ 
0.087 \\
874 & 13 53 03.736 & -28 25 30.935 & 19.759 $\pm$ 0.034 & -14.40 & 1.26 $\pm$ 
0.059 \\
901 & 13 53 03.161 & -28 26 12.292 & 19.776 $\pm$ 0.035 & -14.39 & 2.577 $\pm$ 
0.092 \\
328 & 13 52 44.171 & -28 31 14.701 & 19.797 $\pm$ 0.027 & -14.37 & 2.523 $\pm$ 
0.090 \\
982 & 13 53 16.312 & -28 26 43.636 & 19.805 $\pm$ 0.028 & -14.36 & 0.96 $\pm$ 
0.050 \\
298 & 13 52 41.097 & -28 31 29.559 & 19.941 $\pm$ 0.029 & -14.22 & 0.893 $\pm$ 
0.051 \\
1003 & 13 52 56.896 & -28 27 24.871 & 20.003 $\pm$ 0.031 & -14.16 & 1.235 
$\pm$ 0.061 \\
\hline
\end{tabular}

\medskip 
{\bf $^1$~}{SExtractor object identification}
\end{table*}

\begin{table*}
\caption{Faint  group member candidates. (cont.)}
\label{tab7}
\begin{tabular}{rccccc} \hline
 {object$^1$}   & {$\alpha$} & {$\delta$} & {m$_R$}   & {M$_R$}   & {B - R} \\
                         & (J2000.0) & (J2000.0)& [mag]     & [mag]        & 
[mag]   \\
\hline
76 & 13 52 48.699 & -28 33 16.953 & 20.058 $\pm$ 0.035 & -14.10 & 1.458 $\pm$ 
0.069 \\
739 & 13 53 13.683 & -28 25 00.555 & 20.087 $\pm$ 0.034 & -14.08 & 1.517 $\pm$ 
0.070 \\
607 & 13 53 03.679 & -28 29 12.862 & 20.091 $\pm$ 0.031 & -14.07 & 1.466 $\pm$ 
0.068 \\
141 & 13 52 49.107 & -28 32 46.192 & 20.101 $\pm$ 0.036 & -14.06 & 1.386 $\pm$ 
0.069 \\
635 & 13 52 45.502 & -28 27 07.050 & 20.102 $\pm$ 0.035 & -14.06 & 1.74 $\pm$ 
0.077 \\
1174 & 13 52 54.039 & -28 26 52.981 & 20.144 $\pm$ 0.034 & -14.02 & 2.353 
$\pm$ 0.100 \\
822 & 13 53 02.007 & -28 27 17.293 & 20.210 $\pm$ 0.033 & -13.95 & 1.908 $\pm$ 
0.086 \\
1201 & 13 53 02.989 & -28 28 48.253 & 20.228 $\pm$ 0.040 & -13.93 & 1.997 
$\pm$ 0.093 \\
596 & 13 52 58.807 & -28 29 09.795 & 20.230 $\pm$ 0.041 & -13.93 & 1.439 $\pm$ 
0.077 \\
299 & 13 52 36.507 & -28 31 31.002 & 20.238 $\pm$ 0.034 & -13.92 & 1.211 $\pm$ 
0.066 \\
315 & 13 52 40.791 & -28 31 22.086 & 20.258 $\pm$ 0.035 & -13.90 & 1.125 $\pm$ 
0.065 \\
74 & 13 52 41.258 & -28 33 22.562 & 20.258 $\pm$ 0.034 & -13.90 & 1.703 $\pm$ 
0.080 \\
282 & 13 52 56.080 & -28 31 37.776 & 20.258 $\pm$ 0.034 & -13.90 & 2.172 $\pm$ 
0.099 \\
1114 & 13 52 46.274 & -28 27 59.216 & 20.292 $\pm$ 0.035 & -13.87 & 1.573 
$\pm$ 0.077 \\
802 & 13 52 59.374 & -28 25 14.930 & 20.298 $\pm$ 0.037 & -13.86 & 1.028 $\pm$ 
0.065 \\
754 & 13 52 59.190 & -28 24 46.133 & 20.309 $\pm$ 0.048 & -13.85 & 1.993 $\pm$ 
0.106 \\
867 & 13 53 06.404 & -28 27 50.912 & 20.310 $\pm$ 0.035 & -13.85 & 1.142 $\pm$ 
0.067 \\
918 & 13 53 16.367 & -28 27 38.204 & 20.332 $\pm$ 0.047 & -13.83 & 1.474 $\pm$ 
0.084 \\
1130 & 13 53 11.607 & -28 28 12.106 & 20.340 $\pm$ 0.037 & -13.82 & 1.564 
$\pm$ 0.080 \\
447 & 13 52 44.218 & -28 30 19.272 & 20.366 $\pm$ 0.037 & -13.80 & 1.278 $\pm$ 
0.072 \\
856 & 13 52 48.514 & -28 27 41.771 & 20.430 $\pm$ 0.041 & -13.73 & 2.053 $\pm$ 
0.103 \\
620 & 13 53 14.759 & -28 29 17.774 & 20.435 $\pm$ 0.036 & -13.73 & 1.816 $\pm$ 
0.090 \\
307 & 13 52 42.793 & -28 31 26.273 & 20.452 $\pm$ 0.042 & -13.71 & 1.613 $\pm$ 
0.089 \\
114 & 13 52 53.684 & -28 32 55.839 & 20.480 $\pm$ 0.048 & -13.68 & 1.608 $\pm$ 
0.097 \\
541 & 13 52 50.344 & -28 28 40.114 & 20.486 $\pm$ 0.038 & -13.68 & 0.91 $\pm$ 
0.067 \\
1023 & 13 52 46.371 & -28 26 54.904 & 20.519 $\pm$ 0.049 & -13.64 & 1.307 
$\pm$ 0.084 \\
421 & 13 52 51.490 & -28 30 30.187 & 20.519 $\pm$ 0.039 & -13.64 & 1.803 $\pm$ 
0.094 \\
1129 & 13 52 47.014 & -28 28 10.742 & 20.540 $\pm$ 0.039 & -13.62 & 1.043 
$\pm$ 0.072 \\
200 & 13 52 44.304 & -28 31 58.908 & 20.564 $\pm$ 0.046 & -13.60 & 1.58 $\pm$ 
0.095 \\
1030 & 13 53 10.184 & -28 27 47.998 & 20.626 $\pm$ 0.064 & -13.54 & 1.7 $\pm$ 
0.111 \\
464 & 13 53 03.242 & -28 30 10.125 & 20.667 $\pm$ 0.042 & -13.50 & 1.48 $\pm$ 
0.089 \\
919 & 13 53 15.473 & -28 27 49.968 & 20.695 $\pm$ 0.046 & -13.47 & 1.558 $\pm$ 
0.098 \\
1020 & 13 53 01.946 & -28 27 10.920 & 20.700 $\pm$ 0.045 & -13.46 & 1.278 
$\pm$ 0.086 \\
1187 & 13 53 11.477 & -28 28 36.067 & 20.733 $\pm$ 0.044 & -13.43 & 1.392 
$\pm$ 0.090 \\
1043 & 13 52 53.579 & -28 27 17.626 & 20.774 $\pm$ 0.054 & -13.39 & 1.178 
$\pm$ 0.091 \\
762 & 13 53 08.939 & -28 27 47.274 & 20.828 $\pm$ 0.047 & -13.33 & 0.847 $\pm$ 
0.079 \\
247 & 13 52 47.250 & -28 31 52.194 & 20.963 $\pm$ 0.054 & -13.20 & 1.45 $\pm$ 
0.103 \\
702 & 13 52 38.071 & -28 29 44.466 & 20.981 $\pm$ 0.056 & -13.18 & 1.273 $\pm$ 
0.100 \\
\hline
\end{tabular}

\medskip 
{\bf $^1$~}{SExtractor object identification}
\end{table*}

\label{lastpage}

\end{document}